\pdfoutput=1

\documentclass[a4paper]{article}

\usepackage{fullpage}

\usepackage{microtype}
\usepackage{complexity}                
\usepackage{varioref}                  
\usepackage{hyperref}       
\usepackage{cleveref}                  
\usepackage{booktabs}                  
\usepackage{longtable}
\usepackage{tikz}                      
\usepackage{amsmath}                   
\usepackage{xfrac}                     
\usepackage{siunitx}                   
\usepackage{gnuplot-lua-tikz}          
\usepackage[ruled,vlined]{algorithm2e}
\usepackage{afterpage}
\usepackage[font=normalsize]{caption}

\usepackage{natbib}

\usepackage[ruled]{algorithm2e}

\crefname{algocf}{Algorithm}{Algorithms}
\Crefname{algocf}{Algorithm}{Algorithms}
\crefname{figure}{Figure}{Figures}
\Crefname{figure}{Figure}{Figures}
\crefname{table}{Table}{Tables}
\Crefname{table}{Table}{Tables}

\usetikzlibrary{arrows, shadows, calc, positioning, shapes}
\usetikzlibrary{decorations.pathreplacing, decorations.pathmorphing}

\definecolor{safelightblue}{rgb}{0.65098, 0.807843, 0.890196}
\definecolor{safedarkblue}{rgb}{0.121569, 0.470588, 0.705882}
\definecolor{safelightorange}{rgb}{0.996078, 0.961961, 0.907843}
\definecolor{safedarkorange}{rgb}{0.34902, 0.176471, 0.015686}
\definecolor{safeverydarkblue}{rgb}{0.007843, 0.219608, 0.345098}
\definecolor{safemediumorange}{rgb}{0.74902, 0.505882, 0.490196}
\definecolor{safelightgreen}{rgb}{0.698039, 0.87451, 0.541176}
\definecolor{safedarkgreen}{rgb}{0.2, 0.627451, 0.172549}

\colorlet{gp lt color 0}{safeverydarkblue}
\colorlet{gp lt color 1}{safedarkblue}
\colorlet{gp lt color 2}{safelightblue}
\colorlet{gp lt color 3}{safedarkgreen}
\colorlet{gp lt color 4}{safelightgreen}
\colorlet{gp lt color 5}{black}
\colorlet{gp lt color 6}{safedarkorange}
\colorlet{gp lt color 7}{safemediumorange}

\author{
    Ciaran McCreesh%
    \thanks{This work was supported by the Engineering and Physical Sciences Research Council [grant number EP/K503058/1]}
    \\ \href{mailto:c.mccreesh.1@research.gla.ac.uk}{\nolinkurl{c.mccreesh.1@research.gla.ac.uk}}
    \and
    Patrick Prosser
    \\ \href{mailto:patrick.prosser@glasgow.ac.uk}{\nolinkurl{patrick.prosser@glasgow.ac.uk}}
}

\def\figcaptionfont{\bfseries}

\begin{document}


\title{The Shape of the Search Tree for the Maximum Clique Problem, and the Implications for Parallel Branch and Bound}



%
%

\maketitle

\begin{abstract}
Finding a maximum clique in a given graph is one of the fundamental \NP-hard problems. We compare
two multi-core thread-parallel adaptations of a state-of-the-art branch and bound algorithm for
the maximum clique problem, and provide a novel explanation as to why they are successful. We show
that load balance is sometimes a problem, but that the interaction of parallel search order and the
most likely location of solutions within the search space is often the dominating consideration. We
use this explanation to propose a new low-overhead, scalable work splitting mechanism. Our approach
uses explicit early diversity to avoid strong commitment to the weakest heuristic advice, and late
resplitting for balance. More generally, we argue that for branch and bound, parallel
algorithm design should not be performed independently of the underlying sequential algorithm.
\end{abstract}

\section{Introduction}

When parallelising an algorithm, we seek to divide work into a suitable number of equally sized,
independent subproblems which may be evaluated simultaneously. Sequential branch and bound
algorithms do not lend themselves to this approach, and parallel branch and bound algorithms must
cope with extremely irregular subproblem sizes, speculative evaluation and unpredictable
communication. Nonetheless, parallel branch and bound can be beneficial in practice. Here we look at
some of the choices available when parallelising a branch and bound algorithm. We show that
different work splitting strategies can have a substantial effect upon performance, and
that a careful study of the underlying sequential algorithm can help direct our design.

We look at parallel algorithms for the maximum clique problem, and make use of the extra programming
flexibility offered by multi-core systems to take measurements ``inside the search'' to demonstrate
\emph{why} some implementation choices give better or worse performance on some problems.  We make
three claims, and justify them experimentally. Firstly, we show that static work splitting can lead
to balance problems. Secondly, we show that different search orders and work splitting mechanisms
often have a large effect upon speedups, and that this is due to changes in ``the amount of work
done'', rather than just imbalances and overheads. Thirdly, we show that an understanding of how
ordering heuristics behave can explain why a simple fixed-depth work splitting mechanism usually
does so well, despite occasional balance problems. Using these results, we discuss how we may
improve work balance whilst retaining the search-order benefits of a simpler work splitting
mechanism. We present a novel work splitting mechanism which explicitly diversifies at the top of
search, to offset poor early heuristic advice, and that resplits work later to improve balance. We
show that, as our explanations predict, this gives better and more consistent speedups than other
mechanisms.

\subsection{The Maximum Clique Problem}

\begin{figure}[t]
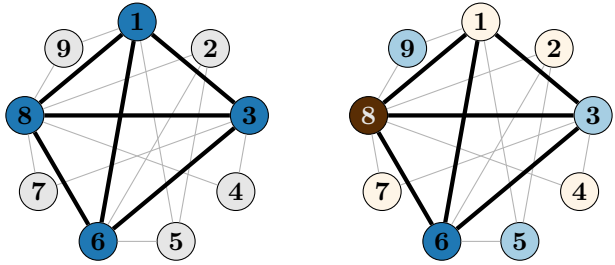

    \centering
    \begin{tikzpicture}
        \input{figure-clique}
    \end{tikzpicture}\hspace{3em}\begin{tikzpicture}
        \input{figure-colour}
    \end{tikzpicture}
    \caption{On the left, a graph, with its unique maximum clique $\{1, 3, 6, 8\}$ of size 4
        highlighted. On the right, the same graph coloured greedily using four colours: vertices
        $\{1, 2, 4, 7\}$ are coloured pale cream, $\{3, 5, 9\}$ are light blue, $\{6\}$ is medium
        blue, and $\{8\}$ is dark chocolate. From this we may infer that there is no clique of size
        larger than 4. We also know that any clique of size 4 must contain vertices $6$ and $8$, plus
        one light blue vertex and one pale cream vertex. This kind of reasoning forms the basis of the
        sequential algorithms.}
    \label{figure:clique}
\end{figure}

A clique in a graph is a set of vertices, each of which is adjacent to every other vertex in this
set. The maximum clique problem is to find the largest such set in a given graph (see
\vref{figure:clique}). This is one of Garey and Johnson's fundamental \NP-hard problems
\citeyearpar{Garey:1990}; applications include bioinformatics \citep{Eblen:2011}, biochemistry
\citep{Butenko:2006}, computer vision, electrical engineering, communications \citep{Bomze:1999} and
controlling flying robots \citep{Regula:2013}. Here we primarily consider dense, computationally
challenging graphs \citep{Cheeseman:1991}.

The starting point for state-of-the-art exact maximum clique solvers is a series of branch and bound
algorithms by Tomita et al.\ \citeyearpar{Tomita:2003,Tomita:2007,Tomita:2010}. San Segundo et al.\
\citeyearpar{SanSegundo:2011,SanSegundo:2011b} observed that a bitset encoding could substantially
speed up the implementation of these algorithms, without affecting the steps taken; a computational
study by \citet{Prosser:2012} analyses variations of these algorithms in depth.
Subsequently, Batsyn, Maslov et al.\ showed that substantial improvements could be had by priming
these algorithms with a good heuristic solution \citep{Batsyn:2013,Maslov:2013}. (Other exact
approaches have been tried, including constraint programming \citep{Regin:2003} and SAT
\citep{Li:2011}; we focus upon dedicated algorithms here.)

Recently the authors implemented a threaded version of one of these algorithms
\citep{McCreesh:2013}.  Independently, Depolli et al.\ did the same \citep{Depolli:2013}. Both
approaches started with a similar sequential algorithm, produced an implementation using C++11
native threads that explored subtrees in parallel, and performed experiments on
similar multi-core hardware. At a glance, both sets of results are similar: good speedups are
achievable in practice (but the speedups achieved vary substantially between problem instances), and
superlinear speedups sometimes happen---in other words, both papers show that adding
parallel tree-search to a strong maximum clique algorithm is worth doing. A close inspection
of both sets of results shows that for some particular problem instances, there are substantial
differences: for example, the authors obtained roughly linear speedups for the DIMACS graph
``MANN\_a45'', whereas Depolli et al.\ achieved a speedup of below 4 regardless of the number of
processor cores available.  We will explain why this happened, and how this may be addressed without
compromising other results.

We begin with a brief recap of sequential maximum clique algorithms. We show how branch and bound
algorithms may be viewed as a tree search process, and how we can treat behaviour ``inside search''
as different kinds of subtrees. We use this to explain the fundamental concepts underlying parallel
branch and bound, and discuss the potential for speedups.

In \cref{section:design} we look in more detail at the effects of design choices in practice. We
show that the reason for Depolli et al.'s speedup limit of 4 for ``MANN\_a45'' is due to poor load
balancing. We then look in more detail at further problems. We will see that that balance does not
severely restrict speedups in most cases, and that steps taken to improve the balance will often
give much worse performance (and not just due to overheads). Our view of branch and bound as a tree,
together with some knowledge of how heuristics behave, let us explain this. We show that diversity
(that is, using parallelism to hedge against weak early heuristic choices), not balance,
is the primary contributing factor to performance in most of the cases we investigate.

Finally, in \cref{section:best} we discuss how to get ``the best of both worlds''. Driven by a
deeper understanding of the search process, we present a new work splitting technique which
explicitly adds diversity early in search, to offset the weakest heuristic advice, followed by a
low-overhead resplitting mechanism to even out balance problems later on in search.

\section{Algorithms for the Maximum Clique Problem} \label{section:algorithms}

Throughout, let $G = (V, E)$ be a graph with vertex set $V$ and edge set $E$, and let $v \in V$. We
may write $V(G)$ for $V$. The \emph{neighbourhood} of $v$, denoted $N(G, v)$, is the set of vertices
adjacent to $v$. The size of a maximum clique is denoted $\omega$. A \emph{colouring} is an
assignment of vertices to colours, such that adjacent vertices are given different colours (as in
\vref{figure:clique}).

We observe that a clique in $G$ either does not contain $v$, or contains only $v$ and vertices
adjacent to $v$. This allows us to build cliques by recursively selecting a vertex which could be
added to a growing clique, and branching on whether or not to include it. We keep track of the
largest clique found so far, which we call the \emph{incumbent}. We use the size of the incumbent as
a bound: if we can show that a growing clique cannot possibly unseat the incumbent, we may abandon
search and backtrack. We do this by constructing a greedy graph colouring---if a subgraph can be
coloured using $k$ colours, it cannot possibly contain a clique of size greater than $k$ (each
vertex in a clique must be given a different colour).

The sequential algorithms underlying both the authors' parallel approach \citep{McCreesh:2013} and
that by \citet{Depolli:2013} are based upon these principles. We give a brief outline of these
algorithms, and refer the reader to the papers by Tomita et al.\
\citeyearpar{Tomita:2003,Tomita:2007,Tomita:2010} and San Segundo et al.\
\citeyearpar{SanSegundo:2011,SanSegundo:2011b} for detailed analysis.
We note that the vertex ordering at the top of search differs slightly between our two approaches,
so runtimes and node counts are not directly comparable. Prosser's computational study
\citeyearpar{Prosser:2012} discusses these choices in more detail; for our purposes, it suffices to
know that initial vertex ordering can make a substantial difference in the size of the search space,
but that there is no ``best'' ordering.

\subsection{Sequential Branch and Bound}

In \vref{algorithm:outline} we give the underlying sequential procedure. The variable $C$ is a
growing clique, and $P$ contains undecided vertices which could potentially be added to $C$. The
best solution found so far is stored in $C_{max}$.  The important part, for this discussion, is that
we have a $\KwSty{for}$ loop, and that the $\FuncSty{expand}$ function calls itself recursively
inside this loop. Some of these recursive calls are then avoided, by using a greedy graph colouring
to prove that the current portion of the search space could not possibly lead to a better solution
than the one in $C_{max}$ being found.

\begin{algorithm}[t]\DontPrintSemicolon
    \FuncSty{expand} (Graph $G$, Set $C$, Set $P$, Set $C_{max}$) \;
    \nl \Begin{
        \nl Construct a greedy colouring of the subgraph induced by $P$ \;
        \nl \For{$v \in P$~\textnormal{(in reverse colour order)}} {
            \nl \If{$|C| + \textnormal{number of colours used to colour}~P > |C_{max}|$} {
                \nl $C \gets C \cup \{v\}$ \tcp*{Add $v$ to growing clique}
                \nl $P' \gets P \cap N(G, v)$ \tcp*{Filter non-adjacent vertices}
                \nl \If{$P' = \emptyset$}{
                    \nl \lIf{$|C| > |C_{max}|$}{$C_{max} \gets C$ \tcp*[f]{Unseat the incumbent}}
                }
                \nl \lElse{$\FuncSty{expand}(G, C, P', C_{max})$ \tcp*[f]{Recurse with $v \in C$}}
                \nl $C \gets C \setminus \{v\}$ \tcp*{Now consider $v$ not in the clique}
                \nl $P \gets P \setminus \{v\}$ \;
            }
        }
    }
    \;
    \nl \FuncSty{maxClique} (Graph $G$) $\rightarrow$ Set \;
    \nl \Begin{
        \nl Permute $G$ to be in non-increasing degree order \;
        \nl $C_{max} \gets \emptyset$ \;
        \nl $\FuncSty{expand}(G, \emptyset, V(G), C_{max})$ \;
        \nl $\KwSty{return}~C_{max}$ \;
    }
    \caption{An outline of a maximum clique algorithm, with many details omitted.}
    \label{algorithm:outline}
\end{algorithm}

\begin{figure}[t]
    \centering
    \begin{tikzpicture}[scale=0.75]
        \input{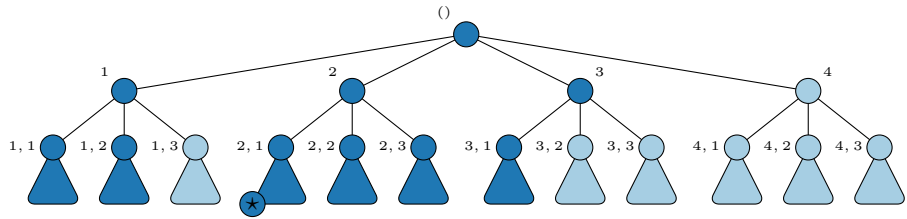}
    \end{tikzpicture}

    \caption{A possible search space of a branch and bound algorithm, viewed as a tree---we assume a
        depth first search, iterating over children from left to right. Each node represents a
        recursive call; triangles represent large subtrees. The optimal solution is marked $\star$,
        at location $(2, 1\ldots)$. Nodes which need not be explored if this optimal solution
        has already been found, which are said to be eliminable, are shaded in light blue; nodes
        which must always be explored to prove optimality are shaded in medium blue. Here, the light blue
        subtree $(1, 3)$ is avoidable---it would be explored in the sequential run despite being
        eliminable.}
    \label{figure:searchspace}
\end{figure}

We may view the recursive calls made by a branch and bound algorithm as forming a tree, as in
\vref{figure:searchspace}. Here, each node in the tree represents a recursive call; triangles
represent large subtrees. We mark by $\star$ the location of an optimal solution. Nodes shown in
medium blue are those which cannot be eliminated by the bound, regardless of the strength of the
incumbent---we say such nodes are \emph{ineliminable}. Nodes in light blue are those which could be
eliminated by the bound, if $\star$ has been found---such nodes are \emph{eliminable}. The tree is
traversed in a manner similar to a depth-first search, exploring subtrees from left to right. Note
that in a sequential run, the leftmost light blue subtree will \emph{not} be eliminated by the
bound, since the search will not yet have found $\star$. In other words, not all eliminable nodes
are necessarily eliminated. We call eliminable nodes that are not eliminated in the sequential run
\emph{avoidable}.

In maximum clique terms, $\star$ is the location of a maximum clique, whose size is denoted
$\omega$.  The medium blue nodes are then the nodes which must be explored to prove that there is no
clique of size $\omega + 1$ in the graph. Light blue nodes are those that would be eliminated by the
bound, if the algorithm were to be initialised with an incumbent size of $\omega$ rather than $0$.
For now we assume there is a unique optimum, and that the optimum must be known before any subtree
may be eliminated by the bound. We will see later that this assumption is slightly too simple in
some cases.

We may label nodes with lists of numbers as follows: the root node is labelled with the empty list,
$()$. We label child of the root node from left to right as $(1)$, $(2)$, $(3)$, etc. The children of
$(1)$ are labelled $(1, 1)$, then $(1, 2)$, $(1, 3)$, and so on. Thus, in our figure, the optimal
solution $\star$ is found at the location $(2, 1\ldots)$, which corresponds to taking the second
branch at the top of search followed by the first branch thereafter.

\subsection{Parallel Branch and Bound, and the Potential for Speedup}

Usually when we wish to parallelise an algorithm, we look for independent units of work which may be
evaluated independently. With branch and bound we cannot take this simple approach, and we must
resort to speculatively executing non-independent subproblems to be able to exploit parallel
hardware. We view branch and bound as a tree search, ignore left-to-right dependencies, and explore
different subtrees in parallel. There is a single incumbent which is shared between every worker.
This technique in general is well known (see, for example, discussions by \citet{Bader:2005} and
\citet{Crainic:2006}). For the maximum clique problem, it was first attempted by
\citet{Pardalos:1998}, where message passing was used on a cluster of four machines, and later, by
the authors \citep{McCreesh:2013} and \citet{Depolli:2013} using threads. Another recent parallel
maximum clique algorithm by \citet{Rossi:2013} does the same, starting with an algorithm designed
for large sparse graphs; again, superlinear speedups were observed. Parallel tree-search was also
used by \citet{Debroni:2011} to close the Keller maximum clique problem, using an algorithm which
exploited special properties of these graphs.

Other approaches have been considered. \citet{Xiang:2013} used MapReduce rather than conventional
parallel branch and bound.  Their emphasis is upon the scalability of their solution: they
demonstrate linear speedups (over a parallel implementation, not a sequential one) on three graphs
from one DIMACS family, and on one further graph. We will see that other families of graphs have
properties which can interfere with this approach. \citet{Bergman:2014} used parallel decision
diagrams, and showed excellent scalability---however, their sequential runtimes were typically much
worse than state of the art algorithms.

With the parallel tree-search approach, we are not dividing a fixed amount of work between processor
cores. Thus we should not always expect to gain a speedup approaching $n$ from $n$ cores (such a
speedup is said to be \emph{linear}). Instead, the speedup obtained will depend not just upon the
algorithm used, but also the nature of the input. There are various possible outcomes, which we
illustrate in \vref{figure:speedupoptions}:

\begin{figure}[tb]
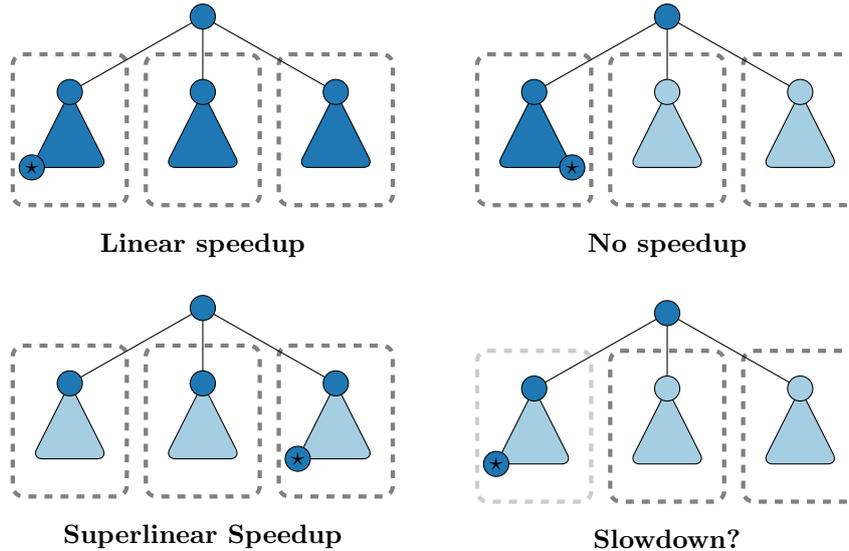

    \centering
    \begin{tikzpicture}
        \input{figure-speedupoptions-linear}
    \end{tikzpicture}\hspace{3em}\begin{tikzpicture}
        \input{figure-speedupoptions-none}
    \end{tikzpicture}

    \vspace{1em}
    \begin{tikzpicture}
        \input{figure-speedupoptions-superlinear}
    \end{tikzpicture}\hspace{3em}\begin{tikzpicture}
        \input{figure-speedupoptions-slowdown}
    \end{tikzpicture}

    \caption{Possibilities for speedup in parallel branch and bound. Here we show the search space
        being divided between three workers. If there are no eliminable nodes, we are dividing up a
        fixed amount of work, and may hope for a linear speedup. If our workers spend their time
        exploring avoidable nodes, as in the top right example, we would get no speedup at all.
        Conversely, a worker may find an optimal solution much more quickly than in the sequential
        run. This may lead to avoidable nodes being eliminated in the parallel run, giving a
        superlinear speedup. Finally, the bottom right example shows the perils of exploring the
        tree ``in a different order'' to the sequential run: if only two workers are available, and
        exploring the leftmost subtree is deferred, we would explore many more nodes in total than in a
        sequential run.}
    \label{figure:speedupoptions}
\end{figure}

\begin{itemize}
    \item The search space consists entirely of nodes that cannot be eliminated---discovering the
        optimal solution does not provide any benefit to proving optimality. In this case, we
        \emph{are} dividing a fixed amount of work up between workers, and may hope for a linear
        speedup.

    \item The search space contains many eliminable nodes that are eliminated on the sequential
        run, but are not eliminated in the parallel run. Our additional workers end up contributing
        nothing to the solution, so we get no speedup at all.

    \item The search space contains many avoidable nodes (eliminable nodes that are not eliminated
        on the sequential run) that are eliminated in parallel due to one of the additional workers
        finding an optimal solution quickly. Here we may get a superlinear speedup
        \citep{Lai:1984,Trienekens:1990}.

    \item We could start by allocating all our workers to explore portions of the search space that
        would be eliminated in the sequential run. This could lead to a slowdown. Intuitively, one
        might think this is due to the tree being explored ``in a different order'' in parallel.
        This possibility is indeed avoidable---we must ensure that incumbents are discovered at
        least as quickly in parallel as they are in the sequential run, and that newly discovered
        incumbents are shared between threads. We refer to works by \citet{Li:1985},
        \citet{Trienekens:1990} and \citet{deBruin:1995} for details (the injectivity
        requirements are not relevant for the maximum clique problem, since all cliques of a given
        size are equivalent for inference purposes).

\end{itemize}

Both our approach and that by Depolli et al.\ meet the conditions for avoiding a slowdown (with one
technicality: it is possible for colourings of a subproblem to be worse than colourings of a parent
problem---we have been unable to observe this having an effect in practice, although it can easily
be worked around \citep{McCreesh:2013.misleading}). The approach by \citet{Xiang:2013} does
\emph{not} meet these conditions: they do not preserve sequential search order, and they do not
share newly discovered incumbents until a subproblem has finished. We will see in the following
section why this was not a problem for the three DIMACS graphs that they considered.

Note that these properties may only be categorised after the fact---when the algorithm finds the
optimal solution, it does not yet know that there is nothing better. Furthermore, in practice, we
would not expect problems to fall neatly into one of these categories---we could both explore some
additional eliminable nodes before a solution is found, and avoid some avoidable nodes.  The matter
is further complicated by the possibility of multiple solutions, and of ``quite good'' solutions
that allow some but not all of the eliminable nodes to be eliminated. Our outcomes only represent
the extremes in behaviour that we might observe.

\section{Do Details of Parallel Algorithm Design Matter?} \label{section:design}

Having discussed what could happen in theory, we now show that these concerns are often relevant in
practice. But first we briefly discuss our experimental setup and test data.

\subsection{Experimental Setup and Data}

Both the authors' previous paper and Depolli et al.\ considered problem instances from the Second
DIMACS Implementation Challenge\footnote{\url{http://dimacs.rutgers.edu/Challenges/}}. We also
considered random graphs and graphs from BHOSLIB (``Benchmarks with Hidden Optimum Solutions for
Graph
Problems'')\footnote{\url{http://www.nlsde.buaa.edu.cn/~kexu/benchmarks/graph-benchmarks.htm}};
Depolli et al.\ also considered protein product
graphs\footnote{\url{http://commsys.ijs.si/~matjaz/maxclique/}}. We will use all three families
here.

We work with a Linux machine with four AMD Opteron 6366 HE processors. Each of these has eight
modules, each containing two cores, for a total of 64 cores. (Note that resources are shared between
pairs of cores in a module, so it is not the case that we have 64 times as much processing power as
is found in a single core used on its own.) The AMD Turbo CORE feature was disabled, to allow us to
investigate scalability effects. We emphasise that each core individually is not especially fast,
compared to the systems previously used by either the authors or Depolli et al.\ in their
experiments---our goal in this paper is not to solve new instances or to solve existing instances
faster, but rather to look at the consequences of algorithm design choices. Speedups are given over
a sequential implementation, not a parallel implementation with one thread. Our compiler is GCC
4.9.0, and we use C++11 native threads.

\subsection{The Importance of Good Work Splitting}

\begin{figure}[t]
    \centering
    \begin{tikzpicture}[scale=0.75]
        \input{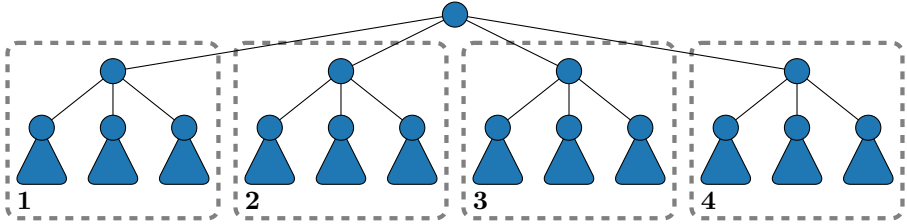}
    \end{tikzpicture}
    \caption{Splitting work at the top: we divide the work into four jobs by splitting the tree at
    distance 1 from the root. Workers tackle the subproblems in the order shown.}
    \label{figure:splitattop}
\end{figure}

We have assumed so far that we are able to divide work between processors; we now discuss how to do
this. A number of approaches are possible. Both the authors and \citet{Depolli:2013} started by
splitting work ``at the top'' of search, as in \vref{figure:splitattop}. Since the number of
vertices in a graph is expected to be much larger than the number of cores available, this produces
more jobs than there are workers. We explicitly placed each subtree of distance 1 from the root onto
a queue (using an additional thread, and blocking when the queue contained too many items), and had
workers process subtrees from this queue in order.  Depolli et al.\ instead started by processing
the root node, and when branching, would process the ``take $v$'' case in one worker, and pass the
``do not take $v$'' case on to a one-item queue to be processed by the next available worker. (We
also had a work donation mechanism for further splitting later on in search; we ignore this for
now.)

\begin{figure}[t]
    \centering
    \begin{tikzpicture}
        \input{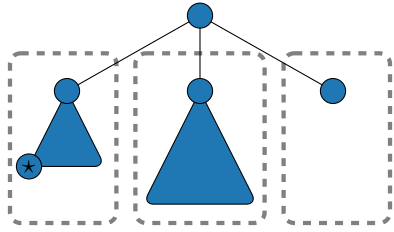}
    \end{tikzpicture}
    \caption{Subtrees can be highly irregularly sized. Here the search space contains no eliminable
    nodes, so if we divided the work between three processor cores as shown we might expect a
    speedup approaching three. But if work is not split dynamically, the runtime may be dominated by
    the cost of exploring the large subtree in the middle.}
    \label{figure:irregular}
\end{figure}

Such an approach assumes that subtrees will be sufficiently numerous and of similar size to allow an
even work distribution. We could imagine a situation like the one shown in
\vref{figure:irregular} occurring, where the cost of evaluating one subtree dominates the
runtime. Informally, we say that a workload is \emph{balanced} if each thread has roughly the same
runtime. We suggest that balance is generally a good thing\footnote{Of course, one could cheat and
achieve a supposedly perfect balance by having threads that would otherwise be idle perform
useless work instead. Thus measures of balance should be used for enlightenment, not for
comparisons.}.

Depolli et al.\ claim their solution offers ``low idle time and good load-balancing'', and for most
problem instances this is indeed the case. However, we will now show that a balance problem is the
reason they were unable to achieve a speedup of 4 for the DIMACS graph ``MANN\_a45'' regardless of
the number of processors used. This contradicts their speculation that the density and large maximum
clique size of the graph was to blame for the atypically poor speedup.

\Vref{figure:graph-togian-MANN_a45-speedup} plots speedups obtained by our implementation as
the number of worker threads increases (speedups are measured over a good sequential implementation,
not a parallel implementation run with one thread). The solid dark blue line shows what happens when
splitting work at distance 1, without the work donation mechanism---we see that our speedups appear
to be capped at around four\footnote{Strictly speaking, it is a coincidence that our limit of
``around~4'' and Depolli et al.'s figure of ``between~3 and~4'' both involve the number~4. The
two sequential algorithms used different initial vertex orderings, and the implementations give
different rates of nodes per second at different depths in the graph, so we should not expect to
see exactly the same limit in both cases.}.

\begin{figure}[p]
    \centering
    \input{gen-graph-togian-MANN_a45-speedup}
    \caption{Speedup obtained as the number of threads increases for the DIMACS instance
        ``MANN\_a45''. We see that with no work donation, splitting at distance 1 limits the speedup
        to around 4 regardless of the number of threads used, and splitting at distance 2 limits the
        speedup to 16. Splitting at distance 3 gives a typical speedup curve. Also shown are results
        for the authors' earlier work donation approach (which evidently has scalability limits at
        this range with this hardware) and a more scalable approach described at the end of this
    paper. The sequential runtime is \SI{438}{\second}.}
    \label{figure:graph-togian-MANN_a45-speedup}

    \vspace{1em}
    \centering
    \input{gen-graph-togian-MANN_a45-idle}
    \caption{The first three graphs show runtimes of individual worker threads as the number of threads
        increases for the DIMACS instance ``MANN\_a45'', with different splitting distances. The top
        line shows the runtime of the longest thread, and the bottom line the shortest. The fourth graph
        shows the runtimes of each thread when using 32 threads. We see that as the splitting
        distance increases, the balance improves.}
    \label{figure:graph-togian-MANN_a45-idle}
\end{figure}

The first graph in \vref{figure:graph-togian-MANN_a45-idle} shows the gap between the
runtime of the shortest running and the longest running threads, as the number of threads increases.
We would expect the gap between the two lines to be small if a good balance has been achieved; here
we see that splitting at distance 1 gives a reasonable balance only for up to four threads.

The dark blue line in the bottom right graph shows runtimes of individual threads, when 32 threads
are used. A perfectly balanced work distribution would give a horizontal line. Here we see the
longest running thread working for \SI{107}{\second}, with the second longest finishing after
\SI{79}{\second}, and there are only six further threads with runtimes over \SI{10}{\second}. The
sequential runtime is \SI{438}{\second}, and our speedup is capped at $\sfrac{438}{107} \approx 4.1$.

\begin{figure}[t]
    \centering
    \begin{tikzpicture}[scale=0.75]
        \input{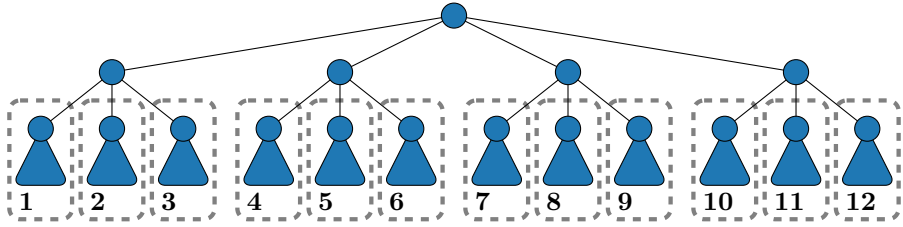}
    \end{tikzpicture}
    \caption{Splitting work at level 2. We might hope that this would lead to a more even work
    distribution than in \cref{figure:splitattop}. But we will see that doing so does not
    avoid the problems with irregularly sized subtrees, and removes one of the benefits of splitting
    work at the top.}
    \label{figure:splitatlevel2}
\end{figure}

One possible workaround is to split the tree further from the root, in the hopes that this will give
a better balance. For example, it is trivial to modify our implementation to split at distance 2
from the root, as in \vref{figure:splitatlevel2}.  The solid medium blue lines in
\vref{figure:graph-togian-MANN_a45-speedup,figure:graph-togian-MANN_a45-idle} show that doing so
solves the balance problem for ``MANN\_a45'' for up to fifteen threads, beyond which again the
runtime is determined by the size of the largest subproblem. Going one step further, and splitting
at distance 3 from the root (shown with solid light blue lines) gives a sufficiently balanced
distribution to allow effective use of 64 cores. This approach is also taken by \citet{Xiang:2013},
with a splitting distance chosen at the start of search based upon estimates of
the sizes of subtrees.

The principle underlying this approach is that we may solve the irregularity problem by breaking the
problem up into many more pieces than we have cores, so that an even balance is obtained
automatically. A similar approach in a constraint programming setting has been used by
\citet{Regin:2013}, with very favourable results on a range of standard constraint programming
problems (but maximum clique was not considered).

We could also consider splitting work entirely dynamically \citep{Sanders:1995}. If the algorithm is
refactored to perform a binary search, this is conceptually simple: before recursing down any
``left'' branch, the implementation can check whether any thread is out of work, and if one is, then
it is given the ``right'' branch to evaluate. More generally, the splitting could be left to a high
level library or language feature---we will investigate this in \cref{section:cilk}, using Intel's
Cilk Plus.  It is important to note that the splitting here is non-deterministic: we will see that
this can lead to large variations in runtimes when solving the same problem instance several times.

Unfortunately, for branch and bound algorithms all of these alternatives have extremely undesirable
effects. Looking ahead to \vref{figure:graph-togian-dimacs-speedup} (which uses the same colours for
lines), we see the opposite of what is shown in \vref{figure:graph-togian-MANN_a45-speedup}: for
other instances, splitting at distance 3 is much worse than splitting at distance 1.  This is
because deeper splitting does not just affect balance; as a side effect, the search order is also
changed.  When splitting at distance 1, we explore the subtrees at $(1)$, $(2)$, $(3)$ and so on, in
that (parallel) order. But when splitting at distance 2, we would instead start by exploring the
subtrees $(1, 1)$, $(1, 2)$, $(1, 3)$ and so on. In a hypothetical situation such as
\vref{figure:searchspace} where a solution is located at $(2, 1, 1\ldots)$, splitting at distance 1
will lead to a solution being found immediately, but splitting at distance 2 will not. In practice,
changing the parallel search order can often have much more significant effects than a potentially
improved balance.  The remainder of this paper explains this in more detail, and shows
how to get close to ``the best of both worlds'' (these are the light and dark green dashed lines in
the graphs).

\subsection{Does Parallel Search Order Matter?}

Different parallel search orders only matter if we are not dividing up a fixed amount of work---that
is, if we are not in the situation shown in the top left of \vref{figure:speedupoptions}. If
there are not many eliminable nodes, exploring the search tree in any order which does not violate
Trienekens' conditions for avoiding a slowdown is acceptable \citep{Trienekens:1990}. We now show
that for many DIMACS instances, there are in fact many \emph{avoidable} search nodes. Thus, even
after ensuring that a slowdown cannot happen, parallel search order is an important consideration.
One clue to this fact is the recent work by Batsyn, Maslov et al.\ which ``primes'' search with a
strong incumbent found via iterated local search \citep{Batsyn:2013,Maslov:2013}. This technique has
yielded substantial improvements for some instances; if there were few avoidable nodes, this would
not happen.

In \vref{table:sequential} we list the DIMACS instances whose sequential runtime is less
than two weeks, then the smaller BHOSLIB instances, and \citet{Depolli:2013}'s protein product graph
instances. We show the size of a maximum clique, $\omega$, and the number of search nodes
(that is, the number of recursive calls made). Next is the number of search nodes that it takes to
prove that the graph does not contain a clique of size $\omega + 1$.  We find this total by
rerunning the algorithm with the size of $C_{max}$ initialised to $\omega$ rather than 0. In other
words, this is the number of nodes in ineliminable subtrees. We then show the proportion of the
search space which consists of avoidable nodes. The final column is discussed below.

\afterpage{
\setlength{\LTcapwidth}{\textwidth}
\small
\begin{longtable}{
        l
        r
        S[table-format = 4e2, table-align-exponent = true, table-text-alignment = right,
        input-decimal-markers = x, input-digits=1234567890., tight-spacing = true]
        S[table-format = 4e2, table-align-exponent = true, table-text-alignment = right,
        input-decimal-markers = x, input-digits=1234567890., tight-spacing = true]
        S[table-format = 3.1]
        l}

        \caption{Properties of the sequential search space for selected problem instances. Shown is
        the size of a maximum clique $\omega$, then the total number of search nodes. Next is
        the number of search nodes required to prove there is no clique of size $\omega + 1$,
        and then the percentage of the search space which is avoidable. Finally, we give the
location of the first maximum clique found.\label{table:sequential}} \\

    \toprule

    \multicolumn{1}{l}{\textbf{Instance}} &
    \multicolumn{1}{c}{\textbf{$\omega$}} &
    \multicolumn{1}{c}{\textbf{Total}} &
    \multicolumn{1}{c}{\textbf{Prove}} &
    \multicolumn{1}{c}{\textbf{Avoid}} &
    \multicolumn{1}{l}{\textbf{Location}} \\
    \midrule

    \endfirsthead

    \toprule

    \multicolumn{1}{l}{\textbf{Instance}} &
    \multicolumn{1}{c}{\textbf{$\omega$}} &
    \multicolumn{1}{c}{\textbf{Total}} &
    \multicolumn{1}{c}{\textbf{Prove}} &
    \multicolumn{1}{c}{\textbf{Avoid}} &
    \multicolumn{1}{l}{\textbf{Location}} \\
    \midrule

    \endhead

    \bottomrule
    \endfoot

    brock200\_1 & 
21
 & 
5.25e+05 & 
3.06e+05 & 41.7
 & 
$22, 4, 10, 6, 1{\times}17$
\\
brock200\_2 & 
12
 & 
3.83e+03 & 
2.58e+03 & 32.6
 & 
$10, 7, 1{\times}10$
\\
brock200\_3 & 
15
 & 
1.46e+04 & 
1.45e+04 & 0.3
 & 
$1, 1, 3, 1{\times}12$
\\
brock200\_4 & 
17
 & 
5.87e+04 & 
3.16e+04 & 46.2
 & 
$36, 20, 5, 1{\times}14$
\\
brock400\_1 & 
27
 & 
1.98e+08 & 
1.17e+08 & 41.0
 & 
$20, 2, 10, 11, 4, 2, 1{\times}21$
\\
brock400\_2 & 
29
 & 
1.46e+08 & 
4.84e+07 & 66.8
 & 
$13, 8, 10, 2, 1, 2, 1{\times}23$
\\
brock400\_3 & 
31
 & 
1.20e+08 & 
1.66e+07 & 86.2
 & 
$14, 10, 1, 4, 2, 1{\times}26$
\\
brock400\_4 & 
33
 & 
5.44e+07 & 
7.67e+06 & 85.9
 & 
$9, 3, 4, 3, 1{\times}29$
\\
brock800\_1 & 
23
 & 
2.23e+09 & 
1.76e+09 & 21.0
 & 
$16, 9, 9, 16, 3, 1{\times}18$
\\
brock800\_2 & 
24
 & 
2.24e+09 & 
1.31e+09 & 41.4
 & 
$23, 65, 1{\times}22$
\\
brock800\_3 & 
25
 & 
2.15e+09 & 
7.03e+08 & 67.3
 & 
$46, 2, 24, 6, 2, 1{\times}20$
\\
brock800\_4 & 
26
 & 
6.40e+08 & 
5.09e+08 & 20.5
 & 
$4, 32, 7, 2, 1{\times}22$
\\
C125.9 & 
34
 & 
5.02e+04 & 
2.69e+04 & 46.4
 & 
$7, 5, 5, 3, 1, 1, 2, 2, 1{\times}26$
\\
C250.9 & 
44
 & 
1.08e+09 & 
9.68e+08 & 10.5
 & 
$4, 32, 4, 9, 3, 5, 2{\times}3, 5,\ldots$
\\
C2000.5 & 
16
 & 
1.82e+10 & 
1.82e+10 & 0.0
 & 
$1, 21, 19, 26, 5, 1{\times}11$
\\
c-fat200-1 & 
12
 & 
24e0
 & 
3e0
 & 87.5
 & 
$5, 1{\times}11$
\\
c-fat200-2 & 
24
 & 
24e0
 & 
1e0
 & 95.8
 & 
$1{\times}24$
\\
c-fat200-5 & 
58
 & 
139e0
 & 
27e0
 & 80.6
 & 
$28, 1{\times}57$
\\
c-fat500-1 & 
14
 & 
14e0
 & 
1e0
 & 92.9
 & 
$1{\times}14$
\\
c-fat500-2 & 
26
 & 
26e0
 & 
1e0
 & 96.2
 & 
$1{\times}26$
\\
c-fat500-5 & 
64
 & 
64e0
 & 
1e0
 & 98.4
 & 
$1{\times}64$
\\
c-fat500-10 & 
126
 & 
126e0
 & 
1e0
 & 99.2
 & 
$1{\times}126$
\\
DSJC500\_5 & 
13
 & 
1.15e+06 & 
1.09e+06 & 5.2
 & 
$8, 38, 3, 1{\times}10$
\\
DSJC1000\_5 & 
15
 & 
7.70e+07 & 
7.67e+07 & 0.3
 & 
$1, 96, 1{\times}13$
\\
gen200\_p0.9\_44 & 
44
 & 
1.77e+06 & 
1.49e+05 & 91.6
 & 
$8, 4, 2, 2, 1{\times}40$
\\
gen200\_p0.9\_55 & 
55
 & 
1.70e+05 & 
2.32e+03 & 98.6
 & 
$4, 3, 1{\times}53$
\\
gen400\_p0.9\_65 & 
65
 & 
1.76e+11 & 
7.29e+09 & 95.9
 & 
$11, 7, 3, 2, 1{\times}61$
\\
gen400\_p0.9\_75 & 
75
 & 
1.05e+11 & 
1.24e+08 & 99.9
 & 
$17, 8, 3, 4, 1{\times}71$
\\
hamming6-2 & 
32
 & 
32e0
 & 
1e0
 & 96.9
 & 
$1{\times}32$
\\
hamming6-4 & 
4
 & 
82e0
 & 
81e0
 & 1.2
 & 
$1{\times}4$
\\
hamming8-2 & 
128
 & 
128e0
 & 
1e0
 & 99.2
 & 
$1{\times}128$
\\
hamming8-4 & 
16
 & 
3.65e+04 & 
3.64e+04 & 0.0
 & 
$1{\times}16$
\\
hamming10-2 & 
512
 & 
512e0
 & 
1e0
 & 99.8
 & 
$1{\times}512$
\\
johnson8-2-4 & 
4
 & 
24e0
 & 
23e0
 & 4.2
 & 
$1{\times}4$
\\
johnson8-4-4 & 
14
 & 
126e0
 & 
115e0
 & 8.7
 & 
$1{\times}14$
\\
johnson16-2-4 & 
8
 & 
2.56e+05 & 
2.56e+05 & 0.0
 & 
$1{\times}8$
\\
keller4 & 
11
 & 
1.37e+04 & 
1.37e+04 & 0.3
 & 
$1, 4, 1, 5, 1{\times}7$
\\
keller5 & 
27
 & 
5.07e+10 & 
5.07e+10 & 0.0
 & 
$1, 2, 1, 13, 1, 1, 3, 1{\times}20$
\\
MANN\_a9 & 
16
 & 
71e0
 & 
60e0
 & 15.5
 & 
$1{\times}16$
\\
MANN\_a27 & 
126
 & 
3.80e+04 & 
3.78e+04 & 0.6
 & 
$1{\times}6, 3, 1{\times}119$
\\
MANN\_a45 & 
345
 & 
2.85e+06 & 
2.85e+06 & 0.2
 & 
$1{\times}5, 6, 1{\times}4, 6, 1{\times}334$
\\
p\_hat300-1 & 
8
 & 
1.48e+03 & 
1.29e+03 & 12.9
 & 
$18, 12, 1{\times}6$
\\
p\_hat300-2 & 
25
 & 
4.26e+03 & 
2.83e+03 & 33.5
 & 
$7, 18, 4, 4, 1{\times}21$
\\
p\_hat300-3 & 
36
 & 
6.25e+05 & 
2.46e+05 & 60.6
 & 
$69, 21, 2, 3, 2{\times}3, 1{\times}5, 2, 1{\times}23$
\\
p\_hat500-1 & 
9
 & 
9.78e+03 & 
9.70e+03 & 0.8
 & 
$3, 18, 1{\times}7$
\\
p\_hat500-2 & 
36
 & 
1.14e+05 & 
3.96e+04 & 65.3
 & 
$114, 10, 10, 5, 2, 4, 1{\times}10, 2, 1{\times}19$
\\
p\_hat500-3 & 
50
 & 
3.93e+07 & 
1.56e+07 & 60.3
 & 
$96, 64, 12, 8, 2{\times}3, 1{\times}4, 2, 1{\times}38$
\\
p\_hat700-1 & 
11
 & 
2.66e+04 & 
1.62e+04 & 39.3
 & 
$259, 4, 2, 1{\times}8$
\\
p\_hat700-2 & 
44
 & 
7.51e+05 & 
3.79e+05 & 49.6
 & 
$62, 35, 18, 6, 2, 3, 1{\times}38$
\\
p\_hat700-3 & 
62
 & 
2.82e+08 & 
1.60e+08 & 43.3
 & 
$147, 62, 42, 11, 6, 8, 2, 2,\ldots$
\\
p\_hat1000-1 & 
10
 & 
1.77e+05 & 
1.75e+05 & 0.8
 & 
$6, 29, 5, 1{\times}7$
\\
p\_hat1000-2 & 
46
 & 
3.45e+07 & 
2.18e+07 & 36.8
 & 
$171, 36, 25, 5, 2, 2, 1, 2,\ldots$
\\
p\_hat1000-3 & 
68
 & 
1.30e+11 & 
3.85e+10 & 70.5
 & 
$368, 30, 99, 6, 10, 3, 2, 1,\ldots$
\\
p\_hat1500-1 & 
12
 & 
1.18e+06 & 
9.59e+05 & 19.1
 & 
$226, 75, 4, 1{\times}9$
\\
p\_hat1500-2 & 
65
 & 
2.01e+09 & 
1.09e+09 & 45.5
 & 
$280, 136, 25, 29, 8, 1, 3, 3,\ldots$
\\
san200\_0.7\_1 & 
30
 & 
1.34e+04 & 
227e0
 & 98.3
 & 
$2{\times}3, 1{\times}27$
\\
san200\_0.7\_2 & 
18
 & 
464e0
 & 
1e0
 & 99.8
 & 
$7, 3, 3, 1{\times}15$
\\
san200\_0.9\_1 & 
70
 & 
8.73e+04 & 
18e0
 & 100.0
 & 
$2, 1{\times}69$
\\
san200\_0.9\_2 & 
60
 & 
2.30e+05 & 
1.06e+03 & 99.5
 & 
$5, 1, 3, 1{\times}57$
\\
san200\_0.9\_3 & 
44
 & 
6.82e+06 & 
4.19e+05 & 93.8
 & 
$3, 4, 2, 1{\times}41$
\\
san400\_0.5\_1 & 
13
 & 
2.45e+03 & 
1e0
 & 100.0
 & 
$20, 5, 3, 1{\times}10$
\\
san400\_0.7\_1 & 
40
 & 
1.19e+05 & 
9.94e+03 & 91.7
 & 
$13, 2, 3, 1{\times}37$
\\
san400\_0.7\_2 & 
30
 & 
8.89e+05 & 
8.32e+04 & 90.6
 & 
$27, 1, 2, 1, 2, 1{\times}25$
\\
san400\_0.7\_3 & 
22
 & 
5.21e+05 & 
6.70e+04 & 87.1
 & 
$3, 4, 2, 1, 1, 2, 2, 1{\times}15$
\\
san400\_0.9\_1 & 
100
 & 
4.54e+06 & 
3.33e+05 & 92.7
 & 
$8, 1{\times}99$
\\
san1000 & 
15
 & 
1.51e+05 & 
1e0
 & 100.0
 & 
$42, 39, 1{\times}13$
\\
sanr200\_0.7 & 
18
 & 
1.53e+05 & 
1.26e+05 & 17.6
 & 
$5, 4, 8, 6, 2, 1{\times}13$
\\
sanr200\_0.9 & 
42
 & 
1.49e+07 & 
1.02e+07 & 31.7
 & 
$8, 15, 2, 4, 2{\times}6, 1{\times}32$
\\
sanr400\_0.5 & 
13
 & 
3.20e+05 & 
1.96e+05 & 38.8
 & 
$95, 12, 2, 2, 1{\times}9$
\\
sanr400\_0.7 & 
21
 & 
6.44e+07 & 
6.40e+07 & 0.7
 & 
$1, 22, 4, 11, 6, 1, 3, 2, 1{\times}13$
\\
\midrule
frb30-15-1 & 
30
 & 
2.92e+08 & 
2.18e+08 & 25.5
 & 
$12, 5, 13, 7, 4, 2, 1{\times}3, 3,\ldots$
\\
frb30-15-2 & 
30
 & 
5.57e+08 & 
3.32e+08 & 40.4
 & 
$20, 13, 7, 4, 11, 1, 3, 2,\ldots$
\\
frb30-15-3 & 
30
 & 
1.67e+08 & 
1.00e+08 & 40.0
 & 
$13, 3, 8, 4, 4, 1, 1, 4,\ldots$
\\
frb30-15-4 & 
30
 & 
9.91e+08 & 
4.20e+08 & 57.7
 & 
$55, 10, 12, 4, 2, 1, 2, 1,\ldots$
\\
frb30-15-5 & 
30
 & 
2.83e+08 & 
1.77e+08 & 37.3
 & 
$9, 17, 7, 4, 3, 2, 2, 1{\times}3,\ldots$
\\
\midrule
1KZKA\_3KT2A\_78 & 
247
 & 
247e0
 & 
1e0
 & 99.6
 & 
$1{\times}247$
\\
1allA\_3dbjC\_41 & 
346
 & 
675e0
 & 
372e0
 & 44.9
 & 
$1{\times}346$
\\
1f82A\_1zb7A\_5 & 
500
 & 
716e0
 & 
294e0
 & 58.9
 & 
$1{\times}500$
\\
2FDVC\_1PO5A\_83 & 
556
 & 
1.35e+03 & 
146e0
 & 89.2
 & 
$2, 1{\times}555$
\\
2UV8I\_2J6IA\_13107 & 
69
 & 
4.26e+03 & 
461e0
 & 89.2
 & 
$5, 3, 1, 1, 3, 2, 1{\times}63$
\\
2W00B\_3H1TA\_10858 & 
143
 & 
7.77e+05 & 
1.23e+05 & 84.2
 & 
$6, 6, 1{\times}141$
\\
2W4JA\_2A2AD\_0 & 
447
 & 
890e0
 & 
10e0
 & 98.9
 & 
$1{\times}5, 2, 1{\times}441$
\\
3HRZA\_2HR0A\_476 & 
563
 & 
9.35e+05 & 
3.23e+05 & 65.4
 & 
$1, 2, 2, 1, 2, 1{\times}4, 2, 1{\times}14,\ldots$
\\
3P0KA\_3GWLB\_0 & 
89
 & 
90e0
 & 
3e0
 & 96.7
 & 
$1{\times}89$
\\
3ZY0D\_3ZY1A\_110 & 
52
 & 
52e0
 & 
1e0
 & 98.1
 & 
$1{\times}52$
 \\
\end{longtable}
}

For some DIMACS instances (e.g.\ ``hamming8-4'', ``johnson16-2-4'', the ``keller'' and ``MANN''
graphs and ``sanr400\_0.7''), we see that there are very few avoidable nodes, so parallel search
order does not matter. We note that for each of these graphs, Batsyn, Maslov et al.\ failed to
obtain a substantial improvement by priming the search (their main sequential algorithm differs
slightly, although this does not seem to have much effect here). The DIMACS graphs considered by
Xiang et al.\ are also in this category.

But for more than half of the DIMACS instances, at least a third of the search space is avoidable.
For ``san400\_0.5\_1'' and ``san1000'', the entire search space is avoidable---that is, the bound at
the top of search is sufficient to prove optimality, if a clique of size $\omega$ has already been
found. For the remainder of the ``san'' family (but not ``sanr''), at least \SI{87}{\percent} of the
search space is avoidable.  The ``brock'', ``gen'' and ``p\_hat'' graph families also contain many
members with substantial avoidable proportions---some members of these families of graphs also gave
extremely large speedups with Batsyn, Maslov et al.'s approach. The BHOSLIB instances are similar:
the avoidable proportion is between \SI{25}{\percent} and \SI{58}{\percent}. Finally, the protein
product graphs all have very high avoidable proportions, although in some cases these are coupled
with a low total node count.  Thus we should expect that parallel search order will often, but not
always, be an important factor in determining speedups.

\subsection{The Quality of Heuristics, and What This Implies}

Harvey and Ginsberg's limited discrepancy search \citeyearpar{Harvey:1995} is a general tree-search
technique which is based upon two principles. Firstly, that when a search fails to find a solution
immediately, it is likely that it only made a small number of ``wrong turns'' (a discrepancy is when
search does not follow a heuristic, in an attempt to correct one of these wrong turns). Secondly,
they claim that ``for many problems the heuristics are \emph{least} reliable early in the search,
before making decisions that reduce the problem to a size for which the heuristics become
reliable''. There is a long-standing tradition of focusing upon the first of these two claims, and
ignoring the second \citep{Korf:1996,Walsh:1997,Prosser:2011,Moisan:2013}.  Here we will buck the
trend and emphasise the second claim.

One observation that suggests that the second claim might hold for the maximum clique problem is
that the heuristics use degree information at the top of search. For many of the DIMACS graphs the
degrees of most vertices are fairly similar. This has lead to sophisticated tie-breaking mechanisms
based upon the sum of neighbouring degrees \citep{Tomita:2007,SanSegundo:2014}; Prosser's
computational study suggests that sometimes these help, and sometimes they make things worse
\citep{Prosser:2012}. We consider the possibility that using parallelism to avoid a strong
commitment to the first choice made by a weak early heuristic may be a more fruitful alternative
than trying to wring even more information out of the graph at the top of search. Explicitly
splitting at distance 1 maximises this effect---by assigning additional workers to go against
initial heuristic advice, we minimise early commitments.

We return to \cref{table:sequential} to justify this. The final column shows
the location in the search space of the maximum clique found by the sequential algorithm. If
heuristics were strong at the top of search, we would expect the first number to be 1 in most cases,
and low in the remainder. Sometimes this is indeed the case---for example, for the graphs
``hamming8-4'', taking the first heuristic choice sixteen times in a row leads immediately to an
optimal solution.  Similarly, the heuristic is perfect for ``johnson16-2-4'' (taking the first
heuristic choice at each level again finds a solution immediately), and for ``MANN\_a27'' and
``MANN\_a45'' it is accurate for the first few levels. For these graphs, the cost is almost entirely
in proving optimality. This suggests that having at least one thread preserve search order to avoid
a slowdown is worthwhile (although in fact for all of these graphs, there are multiple maximum
cliques).

On the other hand, for most graphs the first number in the last column is not a 1, or even a small
number. The table shows that Harvey and Ginsberg's second claim holds for the DIMACS, BHOSLIB and
non-trivial protein instances for the maximum clique problem: heuristic information is weak at the
top of search, and strong commitment to that information can result in reduced performance. Combined
with the high proportion of avoidable nodes in many of these cases, we should expect parallel search
order to matter, and for explicit diversity early in search to be beneficial compared to strong
early commitment. (Additionally, early diversity remains as good as any other solution in cases
where heuristics are strong at all levels, or where there are few avoidable nodes.)

\subsection{Selected Results in Depth}

We will now look in more detail at the behaviour of different work splitting mechanisms for selected
instances, to justify our claims about parallel algorithm design.

\begin{figure}[p]
    \centering
    \input{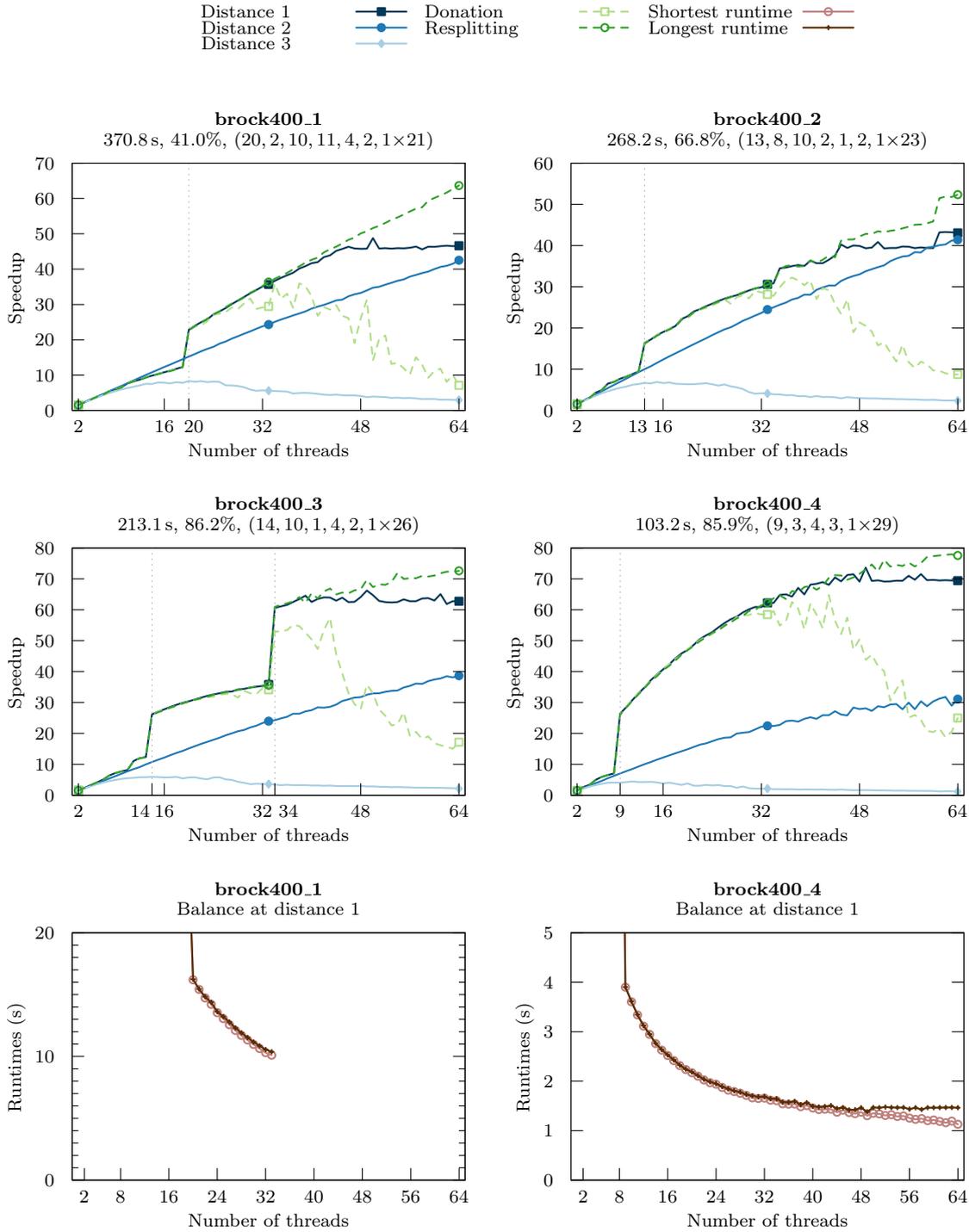}
    \caption{The first four graphs show speedup obtained as the number of threads increases for the
        ``brock400'' family of DIMACS instances.  Measurements are from a 32 module / 64 core
        system. The title shows the instance, the sequential runtime, the proportion
        of avoidable nodes, and the location of the solution found by the sequential algorithm. The
        dotted vertical lines show where we might expect speedup jumps to occur when splitting at
        distance 1; the second dotted vertical line in the third graph at position 34 is explained
        in the text. The bottom two graphs show the balance when splitting at distance 1 with
        varying number of threads for ``brock400\_1'' (where balance becomes a problem) and
        ``brock400\_4'' (where balance is better).}
    \label{figure:graph-togian-dimacs-speedup}
\end{figure}

In \vref{figure:graph-togian-dimacs-speedup} we show the speedups obtained as the number of threads
increases from 2 to 64 for the four members of the DIMACS ``brock400'' family.  We see straight away
that we are getting very different behaviour to that shown for ``MANN\_a45'' in
\vref{figure:graph-togian-MANN_a45-speedup}. For ``MANN\_a45'', splitting at distance 3 was clearly
the best option, but here in each case splitting at distance 1 is best, and distance 3 is very poor.
This shows that balance is not the deciding factor here.

For each of these graphs we see a sudden jump in speedups from being approximately linear
to being visibly superlinear when splitting at distance 1. For ``brock400\_1'', the jump occurs when
going from 19 threads to 20, and the solution is located at $(20, 2, 10, 11, 4, 2, 1{\times}21)$.
Thus with distance 1 splitting, the 20th thread very quickly finds an optimal solution and allows
much of the search tree to be eliminated. The same behaviour occurs for the other three members of
the ``brock400'' family---in each case, the jump occurs with $k$ threads, where $k$ is the first
number in the location of an optimal solution. There is no jump when splitting at distance 2 or
distance 3.

For ``brock400\_3'' ($\omega$ = 31) we get a second jump as we move to 34 threads. We might think
perhaps there is a second optimal solution located at, say, $(34, 1{\times}30)$. In fact it is not
this clear-cut: the solutions for all four of the ``brock400'' graphs are unique, but for
``brock400\_3'' there is a strong (but not optimal) incumbent of size 29 located at $(34, 5, 8, 1,
3, 3, 1{\times}23)$. With 33 threads, the best incumbent found by the equivalent time in the search
only has size 24.

Balance does have some effect for ``brock400\_1'', with over 40 threads---we show this in the fifth
graph. The sixth graph shows ``brock400\_4'', where the balance is even better (the other two graphs
are in-between). But in each case, splitting at distance 1 gives a reasonable work distribution and
best overall performance. We should not find this surprising: Depolli et al.\ did not encounter hard
speedup limits with most instances, so we would not expect balance problems to be common,
particularly with smaller numbers of threads.

One further observation is that in each case, for up to around eight threads it does not matter much
which mechanism is used---each gives a roughly linear speedup.

\begin{figure}[p]
    \centering
    \input{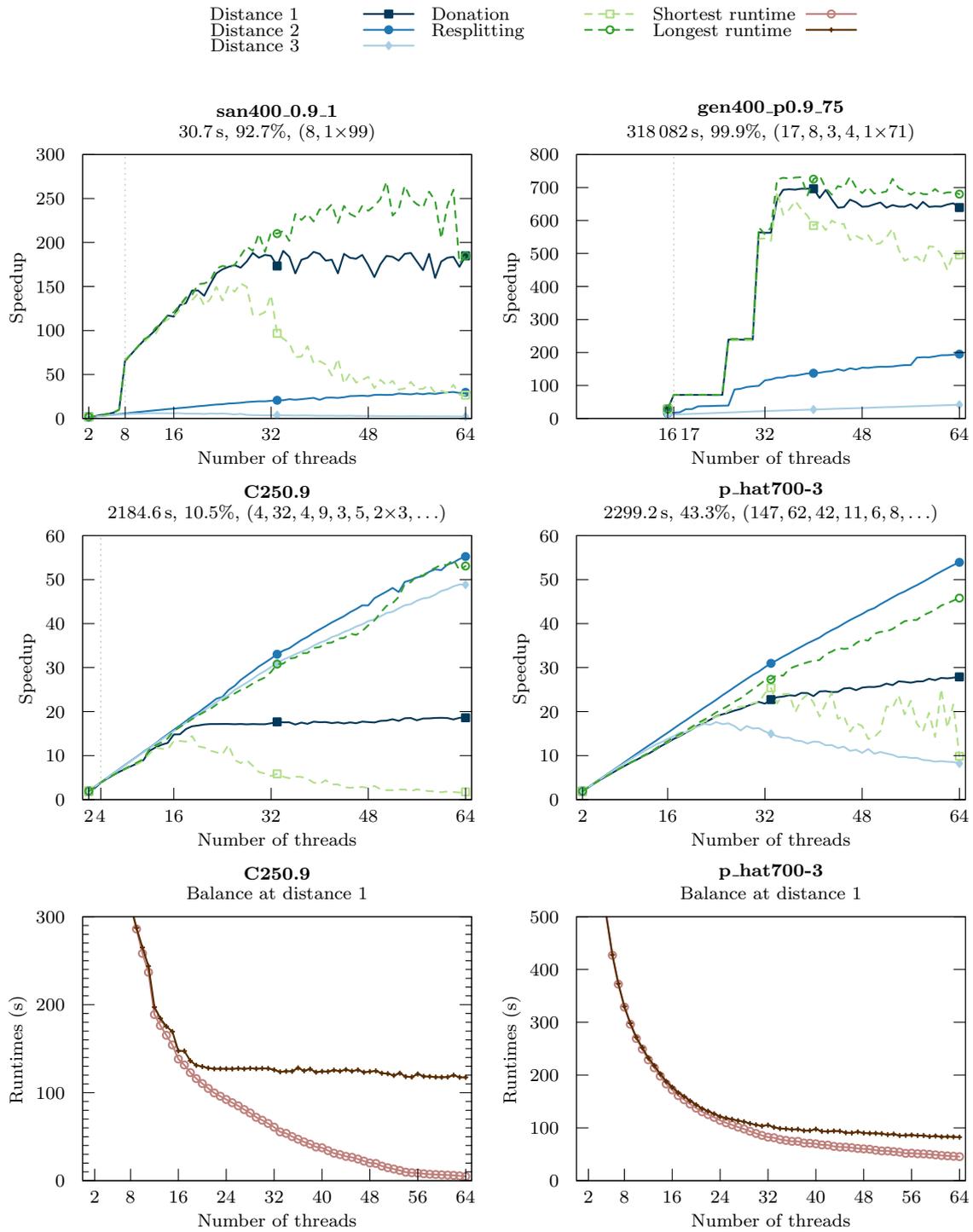}
    \caption{The first four graphs show speedup obtained as the number of threads increases for four
        further DIMACS instances. Measurements are from a 32 module / 64 core system. The title
        shows the instance, the sequential runtime, the proportion of avoidable nodes, and the
        location of the solution found by the sequential algorithm. The dotted vertical lines show
        where we might expect speedup jumps to occur when splitting at distance 1. For
        ``gen400\_p0.9\_75'', results using fewer than 16 threads are omitted due to the long
        sequential runtime. The bottom two graphs show the balance when splitting at distance 1 with
    varying number of threads for ``C250.9'' and ``p\_hat700-3''.}
    \label{figure:graph-togian-dimacs-speedup-2}
\end{figure}

But are these result typical? \Vref{figure:graph-togian-dimacs-speedup-2} shows the same
information for four further DIMACS graphs. Each shows different interesting behaviour.

For ``san400\_0.9\_1'', where the solution is located at $(8, 1{\times}99)$, there is a very sharp
jump as we go from seven to eight threads when splitting at distance 1. Here the eighth thread finds
the solution immediately, and this allows over \SI{90}{\percent} of the search space to be avoided.
We see speedups from nearly 100 to over 300 as we go from 8 to 64 threads. Note that our parallel
runtimes go considerably below one second---within this region, our implementation is sensitive to
scheduling effects, startup costs and the time spent in the initial sequential part of the problem,
and so our speedup lines become rather unstable.

An even stronger superlinear speedup is obtained for ``gen400\_p0.9\_75'', where the speedup from
distance 1 splitting is between 800 and 1000 from 32 or more threads. We should expect that this
could happen: \SI{99.9}{\percent} of the search space is avoidable. But although there is a small
jump when going to 17 threads, as we might predict, there are two much larger jumps at 26 and 31
threads. The jump at 17 threads is small because the heuristic is badly wrong at both the top of
search, and the second level, so the 17th thread still takes a long time to find a maximum clique.
On the other hand, the 26th thread quickly finds an incumbent of size 55 at $(26, 1{\times}3,
2{\times}3, 1, 2, 1{\times}46)$, and a little later the 31st thread finds an incumbent of size 59 at
$(31, 1, 3, 4, 1{\times}3, 2, 1{\times}51)$.  These are followed by larger incumbents being found by
the 17th and 26th threads, and then the 31st thread finds an incumbent of size 72 at $(31, 4, 3,
1{\times}4, 2, 1{\times}64)$. This eliminates enough of the search tree that the maximum clique of
size 75 is then found by the 17th thread as we would expect, and search finishes two seconds later.
This is an unusually complicated picture, whose behaviour is not captured by simple measurements
which assume a single maximum clique and nothing else---we must look more carefully inside the
search process to explain what is going on.

This is not the only peculiar behaviour visible in the graph. We can also explain why the speedup
decreases slightly as the number of threads goes from 32 to 64. Firstly, we note that there is
almost no improvement to the number of nodes required to find an optimal solution when going over 31
threads, so each additional thread could contribute at most a linear extra improvement to the
runtime.  Secondly, we remind the reader that although our hardware is marketed as having 64 cores,
some resources are shared between cores---thus, it does not have 64 times as much processing power
when used perfectly in parallel. The result is that although we have more \emph{total} processing
power when using all 64 cores, each individual core is penalised somewhat: with 32 threads, stronger
incumbents are found slightly sooner than with 64 threads. Since time to find a maximum clique is
the most important factor for this instance, the overall speedup is reduced. This is a slightly odd
case of what \citet{deBruin:1995} describe as the ``[danger of increasing] the processing power of a
system by adding a less powerful processing element''; this should also serve as a warning against
attempting to gain the benefits of increased diversity by using more threads than there are cores.

For ``C250.9'' there is no jump. The reasons for this are twofold. Firstly, although the solution is
found by taking the fourth heuristic choice at the top of search, the second heuristic decision
(where we are not explicitly diversifying) is also poor. Thus, even with distance 1 splitting, we do
not find a better incumbent quickly. Secondly, the proportion of avoidable nodes for this instance
is relatively low, so finding a stronger incumbent quickly does not provide much benefit. This graph
also has a speedup limit from imbalances when splitting at distance 1, which we show in the fifth
graph.

Nor is there a jump for ``p\_hat700-3'' ($\omega$ = 62). Here splitting at distance 2 beats
splitting at distance 1 by a small amount at 16 threads, and a much larger amount by 64 threads.
Balance is one factor here, but only from 24 threads onwards. Splitting at distance 1 finds
incumbents of sizes up to 53 most quickly, but splitting at distance 2 finds incumbents of sizes 54
and higher in slightly less time. Although \SI{43.3}{\percent} of the search space is avoidable, the
heuristics in this case are sufficiently poor at the first six levels that we are unable to find
an optimal solution quickly regardless of how the work is split.

Overall, we see that balance is sometimes a problem with larger numbers of cores, but that
increasing the likelihood of finding an optimal solution quickly is usually far more important. As
we predicted, the early diversity offered by splitting at distance 1 is often helpful with this.

\section{Getting the Best of Both Worlds} \label{section:best}

In view of the previous section, we should look for a work splitting mechanism which gives a good
balance, particularly if we are targeting larger numbers of cores. But the results clearly reinforce
that splitting at distance 2 or 3 rather than distance 1 has practical consequences beyond the
balance of subproblems, and that in most cases parallel search order rather than balance is the
dominating factor. We need an approach which gives the best of both worlds.

It is generally known that parallel search order is important if a strong incumbent is not available
at the start of search, but selecting a parallel search order explicitly designed to improve our
chances of finding a solution quickly has not been given the attention it deserves.  For example,
\citet{Clausen:1997} notes that ``most implementations \ldots focus on workload distribution methods
to increase efficiency of the parallel algorithm''. \citeauthor{Clausen:1997} discusses balance, and
how to assess the efficiency of a parallel algorithm, and observes that ``different selection
strategies may lead to different search trees'', but does not consider where the solutions are
actually likely to be in the search space, and how this can be used to determine a work splitting
mechanism.  Similarly, in the design of an algorithmic skeleton for parallel branch and bound,
\citet{Poldner:2008} state that

\begin{quote}``the number of problems considered by the parallel skeleton differs enormously over
    several runs with the same inputs. This number largely depends on the fact whether a subproblem
    leading to the optimal solution is picked up early or late. Note that the parallel algorithm
    behaves non-deterministically in the way the search-space tree is explored.  In order to get
    reliable results, we have repeated each run 100 times and computed the average runtimes.''
\end{quote}

The search-order issue also occurs in constraint programming. \citet{Caniou:2011} note that
``solutions may be not uniformly distributed in the search space'', and that this has an affect on
parallelism, but they do not consider adapting the search process to improve the chances of finding
a solution quickly. The approach by \citet{Regin:2013} ``relies on the assumption that the
resolution time of disjoint subproblems is equivalent to the resolution time of the union of these
subproblems''---we have seen that this assumption does not hold here.

When proposing a new work splitting mechanism, \citet{Fischetti:2014} state that ``since we are
interested in measuring the scalability of our method, we considered only instances which are either
infeasible or in which we are required to find all feasible solutions (the parallel speedup for
finding a first feasible solution can be completely uncorrelated to the number of workers, making
the results hard to analyze).'' \citet{Leroy:2014} do the same in a branch and bound setting:
``therefore, we chose to always initialize our B\&B by the optimal solution of the instance to be
solved. With this initialization, we are sure that the number of explored subproblems is the same in
both approaches.'' We believe that both of these approaches to measurement and comparison are
flawed---parallelism and search order are linked, and should be considered together. In particular,
we disagree with \citet{RaywardSmith:1993}'s assertion that speedup is ``only a very crude measure
of the success of parallelism'' and their notion of ``pseudo efficiency''. Changes in the amount of
work done are inherent to parallelism, and should \emph{not} be disregarded when evaluating a
parallelism strategy.

We also desire a stronger notion of reliability than \citeauthor{Poldner:2008}. For non-trivial
problem instances, when splitting at a fixed depth, runtimes are very consistent between executions;
we consider it desirable to preserve this property. \citet{Langer:2013} look at parallel branch and
bound for integer programming, and explain that ``reducing idle time by eagerly exploring as much of
the tree as possible might be counter-productive by using compute resources for exploring sub-trees
that might have been easily pruned later''. They evaluate two designs for work allocation, and
observe that ``even though Design A has better repeatability, the worst performance using Design B
is better than the best performance using Design A'', and conclude that ``Design B is the design of
choice''. We do not wish to have to make this kind of trade-off.

\subsection{A Low-Overhead, High-Diversity Work Splitting Mechanism}

We must ensure early diversity (that is, weak heuristic commitment at the top of search), but retain
a way of rebalancing subproblems. Our previous paper described (without detailed justification) a
work donation mechanism, whereby subproblems are requeued when threads become idle; this is the
light green dashed line in our graphs. This mechanism worked very well on the 12 core Intel machine
used in previous experiments, but it is clear that the approach is unsuitable for a 64 core AMD
machine. We now introduce an alternative work splitting mechanism which is more suitable for larger
numbers of cores.

Initially we split work at distance 1, as before, and place items in order onto a queue. When the
queue is empty, and a thread first becomes idle, this thread then steals and requeues unstarted work
from every other thread, splitting at a distance of 2. Finally, when the queue again becomes empty
and a thread becomes idle, work is stolen and requeued with splitting at a distance of 3. This gives
us all the search order benefits of distance 1 splitting, and the balance benefits of splitting at
distance 3. In addition, if $w$ is the number of workers and $|V|$ the number of vertices in the
graph, it limits the queue size to be at worst $w \times |V|$ rather than $|V|^3$, and avoids
enqueueing lots of eliminable nodes early on.

This scheme may be implemented with very little overhead, as follows. Each thread publishes three
integers, describing its current position in search at the first three levels, and three
corresponding flags.  When work stealing takes place, the appropriate flag is set; when returning
from a recursive call at the first three levels, the flag is checked to see whether the remainder of
the work has already been stolen and requeued. In particular, this approach introduces no overhead
below depth 3 (where most of the search time is spent), and is very unlikely to have any contended
critical sections.

Results for this approach are shown using the dark green dashed line in the graphs---we see that in
each case it is at least nearly as good as whichever other mechanism is best, and in many cases it
is slightly better than the best alternative. (Since speedups are being given over a good sequential
implementation, not over a parallel implementation run with one thread, it is legitimate to compare
lines directly.) For all of the ``brock400'' graphs, ``san400\_0.9\_1'' and ``gen400\_p0.9\_75'', we
get the same superlinear speedup jumps as when splitting at distance 1. But ``brock400\_1'',
``MANN\_a45'' and ``C250.9'' show that this approach is successful in addressing imbalances. With
``p\_hat700-3'', we initially behave similarly to splitting at distance 1, but as the number of
threads increases and resplitting starts to have an effect earlier in search, our behaviour
approaches that of distance 2 splitting. The graphs also clearly show that this technique scales
well to 64 cores, and does not introduce noticeable overheads compared to static splitting
mechanisms.

These experiments also suggest that with 64 cores, splitting at distance 3 is sufficient, and there
is no benefit of continuing this to an arbitrary distance. However, this is not an intrinsic
limitation, and a larger maximum depth can be used if necessary for larger numbers of cores or if
unusually-shaped problem instances are found. \citet{Regin:2013} suggest that the number of
subproblems created only needs to grow linearly with the number of cores---increasing the effective
splitting depth by resplitting gives a polynomial increase for linear cost.

\subsection{Comparison to Off-the-Shelf Work Stealing} \label{section:cilk}

\begin{table}[p]
    \setlength{\tabcolsep}{2pt}
    \caption{A comparison of sequential and 64 core parallel runtimes for larger problem instances,
        using either resplitting or Cilk Plus for work stealing. On the first row we show runtimes,
        and on the second, search nodes. Parallel figures are an average of ten samples. The error
        is the standard deviation, and the figures in parentheses are the range, both expressed as a
    proportion. A $\star$ indicates the best average.}
    \label{table:parallel}

        \centering
        \small
        \begin{tabular}{
            l@{\hspace{12pt}}
            r@{\hspace{12pt}}
            rrr@{\hspace{0pt}}r@{\hspace{12pt}}
            rrr@{\hspace{0pt}}r}
        \toprule

        \multicolumn{1}{l}{\textbf{Instance}} &
        \multicolumn{1}{l}{\textbf{Base}} &
        \multicolumn{4}{c}{\textbf{Resplitting}} &
        \multicolumn{4}{c}{\textbf{Cilk Plus}} \\

        \midrule

        brock400\_1 & 
$\SI{370.8}{\second}$
&
$\SI{4.6}{\second}
$
&
$\pm 0.01$
&
$(0.99\text{--}1.02)$
& $\star$
&
$\SI{5.8}{\second}
$
&
$\pm 0.03$
&
$(0.96\text{--}1.04)$
&
\\
&
$\num[tight-spacing=true]{1.98e+08}$
&
$\num[tight-spacing=true]{1.33e+08}$
&
$\pm 0.00$
&
$(1.00\text{--}1.00)$
& $\star$
&
$\num[tight-spacing=true]{1.58e+08}$
&
$\pm 0.04$
&
$(0.95\text{--}1.05)$
&
\\[0.02cm]
brock400\_2 & 
$\SI{268.2}{\second}$
&
$\SI{4.0}{\second}
$
&
$\pm 0.00$
&
$(0.99\text{--}1.01)$
&
&
$\SI{3.5}{\second}
$
&
$\pm 0.05$
&
$(0.92\text{--}1.06)$
& $\star$
\\
&
$\num[tight-spacing=true]{1.46e+08}$
&
$\num[tight-spacing=true]{1.19e+08}$
&
$\pm 0.01$
&
$(0.99\text{--}1.01)$
&
&
$\num[tight-spacing=true]{9.05e+07}$
&
$\pm 0.05$
&
$(0.92\text{--}1.07)$
& $\star$
\\[0.02cm]
brock400\_3 & 
$\SI{213.1}{\second}$
&
$\SI{2.3}{\second}
$
&
$\pm 0.01$
&
$(0.99\text{--}1.01)$
& $\star$
&
$\SI{2.5}{\second}
$
&
$\pm 0.11$
&
$(0.87\text{--}1.22)$
&
\\
&
$\num[tight-spacing=true]{1.20e+08}$
&
$\num[tight-spacing=true]{6.87e+07}$
&
$\pm 0.01$
&
$(0.99\text{--}1.01)$
&
&
$\num[tight-spacing=true]{6.55e+07}$
&
$\pm 0.12$
&
$(0.87\text{--}1.25)$
& $\star$
\\[0.02cm]
brock400\_4 & 
$\SI{103.2}{\second}$
&
$\SI{1.1}{\second}
$
&
$\pm 0.07$
&
$(0.95\text{--}1.19)$
&
&
$\SI{795}{\milli\second}
$
&
$\pm 0.14$
&
$(0.81\text{--}1.27)$
& $\star$
\\
&
$\num[tight-spacing=true]{5.44e+07}$
&
$\num[tight-spacing=true]{2.80e+07}$
&
$\pm 0.05$
&
$(0.93\text{--}1.13)$
&
&
$\num[tight-spacing=true]{1.34e+07}$
&
$\pm 0.23$
&
$(0.75\text{--}1.43)$
& $\star$
\\[0.02cm]
brock800\_1 & 
$\SI{6901.1}{\second}$
&
$\SI{116.7}{\second}
$
&
$\pm 0.00$
&
$(1.00\text{--}1.00)$
& $\star$
&
$\SI{122.7}{\second}
$
&
$\pm 0.01$
&
$(0.98\text{--}1.02)$
&
\\
&
$\num[tight-spacing=true]{2.23e+09}$
&
$\num[tight-spacing=true]{1.95e+09}$
&
$\pm 0.00$
&
$(1.00\text{--}1.00)$
&
&
$\num[tight-spacing=true]{1.92e+09}$
&
$\pm 0.02$
&
$(0.98\text{--}1.03)$
& $\star$
\\[0.02cm]
brock800\_2 & 
$\SI{6890.8}{\second}$
&
$\SI{102.8}{\second}
$
&
$\pm 0.00$
&
$(1.00\text{--}1.00)$
& $\star$
&
$\SI{117.3}{\second}
$
&
$\pm 0.01$
&
$(0.98\text{--}1.02)$
&
\\
&
$\num[tight-spacing=true]{2.24e+09}$
&
$\num[tight-spacing=true]{1.64e+09}$
&
$\pm 0.00$
&
$(1.00\text{--}1.00)$
& $\star$
&
$\num[tight-spacing=true]{1.80e+09}$
&
$\pm 0.02$
&
$(0.98\text{--}1.03)$
&
\\[0.02cm]
brock800\_3 & 
$\SI{6369.0}{\second}$
&
$\SI{52.4}{\second}
$
&
$\pm 0.00$
&
$(1.00\text{--}1.01)$
& $\star$
&
$\SI{83.8}{\second}
$
&
$\pm 0.06$
&
$(0.94\text{--}1.13)$
&
\\
&
$\num[tight-spacing=true]{2.15e+09}$
&
$\num[tight-spacing=true]{7.58e+08}$
&
$\pm 0.00$
&
$(1.00\text{--}1.01)$
& $\star$
&
$\num[tight-spacing=true]{1.30e+09}$
&
$\pm 0.07$
&
$(0.92\text{--}1.16)$
&
\\[0.02cm]
brock800\_4 & 
$\SI{2413.2}{\second}$
&
$\SI{48.6}{\second}
$
&
$\pm 0.00$
&
$(1.00\text{--}1.00)$
& $\star$
&
$\SI{56.0}{\second}
$
&
$\pm 0.06$
&
$(0.88\text{--}1.11)$
&
\\
&
$\num[tight-spacing=true]{6.40e+08}$
&
$\num[tight-spacing=true]{7.15e+08}$
&
$\pm 0.00$
&
$(1.00\text{--}1.00)$
& $\star$
&
$\num[tight-spacing=true]{8.03e+08}$
&
$\pm 0.08$
&
$(0.85\text{--}1.15)$
&
\\[0.02cm]
C250.9 & 
$\SI{2184.6}{\second}$
&
$\SI{41.5}{\second}
$
&
$\pm 0.04$
&
$(0.95\text{--}1.08)$
& $\star$
&
$\SI{43.8}{\second}
$
&
$\pm 0.07$
&
$(0.89\text{--}1.08)$
&
\\
&
$\num[tight-spacing=true]{1.08e+09}$
&
$\num[tight-spacing=true]{1.17e+09}$
&
$\pm 0.05$
&
$(0.95\text{--}1.09)$
& $\star$
&
$\num[tight-spacing=true]{1.17e+09}$
&
$\pm 0.07$
&
$(0.88\text{--}1.09)$
&
\\[0.02cm]
C2000.5 & 
$\SI{33.9}{\hour}$
&
$\SI{2285.7}{\second}
$
&
$\pm 0.00$
&
$(1.00\text{--}1.01)$
&
&
$\SI{2228.8}{\second}
$
&
$\pm 0.00$
&
$(1.00\text{--}1.00)$
& $\star$
\\
&
$\num[tight-spacing=true]{1.82e+10}$
&
$\num[tight-spacing=true]{1.82e+10}$
&
$\pm 0.00$
&
$(1.00\text{--}1.00)$
& $\star$
&
$\num[tight-spacing=true]{1.82e+10}$
&
$\pm 0.00$
&
$(1.00\text{--}1.00)$
&
\\[0.02cm]
DSJC1000\_5 & 
$\SI{181.3}{\second}$
&
$\SI{3.6}{\second}
$
&
$\pm 0.00$
&
$(0.99\text{--}1.01)$
& $\star$
&
$\SI{4.0}{\second}
$
&
$\pm 0.02$
&
$(0.98\text{--}1.04)$
&
\\
&
$\num[tight-spacing=true]{7.70e+07}$
&
$\num[tight-spacing=true]{7.70e+07}$
&
$\pm 0.00$
&
$(1.00\text{--}1.00)$
& $\star$
&
$\num[tight-spacing=true]{7.83e+07}$
&
$\pm 0.00$
&
$(1.00\text{--}1.00)$
&
\\[0.02cm]
gen400\_p0.9\_65 & 
$\SI{156.0}{\hour}$
&
$\SI{4452.2}{\second}
$
&
$\pm 0.00$
&
$(1.00\text{--}1.00)$
& $\star$
&
$\SI{6050.4}{\second}
$
&
$\pm 0.41$
&
$(0.57\text{--}2.04)$
&
\\
&
$\num[tight-spacing=true]{1.76e+11}$
&
$\num[tight-spacing=true]{7.16e+10}$
&
$\pm 0.00$
&
$(1.00\text{--}1.00)$
& $\star$
&
$\num[tight-spacing=true]{1.01e+11}$
&
$\pm 0.44$
&
$(0.57\text{--}2.12)$
&
\\[0.02cm]
gen400\_p0.9\_75 & 
$\SI{88.4}{\hour}$
&
$\SI{348.4}{\second}
$
&
$\pm 0.00$
&
$(1.00\text{--}1.01)$
& $\star$
&
$\SI{3532.5}{\second}
$
&
$\pm 0.35$
&
$(0.47\text{--}1.40)$
&
\\
&
$\num[tight-spacing=true]{1.05e+11}$
&
$\num[tight-spacing=true]{5.51e+09}$
&
$\pm 0.00$
&
$(0.99\text{--}1.01)$
& $\star$
&
$\num[tight-spacing=true]{6.01e+10}$
&
$\pm 0.35$
&
$(0.49\text{--}1.42)$
&
\\[0.02cm]
keller5 & 
$\SI{60.4}{\hour}$
&
$\SI{4022.3}{\second}
$
&
$\pm 0.00$
&
$(1.00\text{--}1.00)$
& $\star$
&
$\SI{4232.1}{\second}
$
&
$\pm 0.00$
&
$(1.00\text{--}1.00)$
&
\\
&
$\num[tight-spacing=true]{5.07e+10}$
&
$\num[tight-spacing=true]{5.07e+10}$
&
$\pm 0.00$
&
$(1.00\text{--}1.00)$
& $\star$
&
$\num[tight-spacing=true]{5.07e+10}$
&
$\pm 0.00$
&
$(1.00\text{--}1.00)$
&
\\[0.02cm]
p\_hat500-3 & 
$\SI{186.2}{\second}$
&
$\SI{2.8}{\second}
$
&
$\pm 0.01$
&
$(0.98\text{--}1.01)$
& $\star$
&
$\SI{3.7}{\second}
$
&
$\pm 0.04$
&
$(0.95\text{--}1.09)$
&
\\
&
$\num[tight-spacing=true]{3.93e+07}$
&
$\num[tight-spacing=true]{3.30e+07}$
&
$\pm 0.01$
&
$(0.99\text{--}1.01)$
& $\star$
&
$\num[tight-spacing=true]{4.26e+07}$
&
$\pm 0.04$
&
$(0.96\text{--}1.08)$
&
\\[0.02cm]
p\_hat700-3 & 
$\SI{2299.2}{\second}$
&
$\SI{49.2}{\second}
$
&
$\pm 0.00$
&
$(1.00\text{--}1.00)$
& $\star$
&
$\SI{51.5}{\second}
$
&
$\pm 0.02$
&
$(0.96\text{--}1.03)$
&
\\
&
$\num[tight-spacing=true]{2.82e+08}$
&
$\num[tight-spacing=true]{3.33e+08}$
&
$\pm 0.00$
&
$(1.00\text{--}1.00)$
&
&
$\num[tight-spacing=true]{3.33e+08}$
&
$\pm 0.02$
&
$(0.96\text{--}1.03)$
& $\star$
\\[0.02cm]
p\_hat1000-2 & 
$\SI{223.9}{\second}$
&
$\SI{5.1}{\second}
$
&
$\pm 0.01$
&
$(0.98\text{--}1.01)$
&
&
$\SI{4.8}{\second}
$
&
$\pm 0.02$
&
$(0.97\text{--}1.03)$
& $\star$
\\
&
$\num[tight-spacing=true]{3.45e+07}$
&
$\num[tight-spacing=true]{4.10e+07}$
&
$\pm 0.00$
&
$(1.00\text{--}1.01)$
&
&
$\num[tight-spacing=true]{3.69e+07}$
&
$\pm 0.02$
&
$(0.98\text{--}1.04)$
& $\star$
\\[0.02cm]
p\_hat1000-3 & 
$\SI{323.8}{\hour}$
&
$\SI{5.7}{\hour}
$
&
$\pm 0.00$
&
$(1.00\text{--}1.00)$
& $\star$
&
$\SI{6.3}{\hour}
$
&
$\pm 0.01$
&
$(0.98\text{--}1.01)$
&
\\
&
$\num[tight-spacing=true]{1.30e+11}$
&
$\num[tight-spacing=true]{1.28e+11}$
&
$\pm 0.00$
&
$(1.00\text{--}1.00)$
& $\star$
&
$\num[tight-spacing=true]{1.37e+11}$
&
$\pm 0.01$
&
$(0.98\text{--}1.02)$
&
\\[0.02cm]
p\_hat1500-2 & 
$\SI{14.0}{\hour}$
&
$\SI{1144.2}{\second}
$
&
$\pm 0.00$
&
$(1.00\text{--}1.00)$
&
&
$\SI{972.7}{\second}
$
&
$\pm 0.02$
&
$(0.97\text{--}1.02)$
& $\star$
\\
&
$\num[tight-spacing=true]{2.01e+09}$
&
$\num[tight-spacing=true]{2.54e+09}$
&
$\pm 0.00$
&
$(1.00\text{--}1.00)$
&
&
$\num[tight-spacing=true]{2.24e+09}$
&
$\pm 0.02$
&
$(0.97\text{--}1.03)$
& $\star$
\\[0.02cm]
sanr200\_0.9 & 
$\SI{27.5}{\second}$
&
$\SI{548}{\milli\second}
$
&
$\pm 0.04$
&
$(0.92\text{--}1.05)$
& $\star$
&
$\SI{865}{\milli\second}
$
&
$\pm 0.09$
&
$(0.90\text{--}1.11)$
&
\\
&
$\num[tight-spacing=true]{1.49e+07}$
&
$\num[tight-spacing=true]{1.43e+07}$
&
$\pm 0.04$
&
$(0.92\text{--}1.05)$
& $\star$
&
$\num[tight-spacing=true]{1.93e+07}$
&
$\pm 0.11$
&
$(0.84\text{--}1.18)$
&
\\[0.02cm]
sanr400\_0.7 & 
$\SI{96.1}{\second}$
&
$\SI{1.8}{\second}
$
&
$\pm 0.01$
&
$(0.99\text{--}1.01)$
& $\star$
&
$\SI{2.1}{\second}
$
&
$\pm 0.02$
&
$(0.98\text{--}1.06)$
&
\\
&
$\num[tight-spacing=true]{6.44e+07}$
&
$\num[tight-spacing=true]{6.54e+07}$
&
$\pm 0.00$
&
$(1.00\text{--}1.00)$
&
&
$\num[tight-spacing=true]{6.47e+07}$
&
$\pm 0.01$
&
$(0.99\text{--}1.01)$
& $\star$
\\[0.02cm]
frb30-15-1 & 
$\SI{999.6}{\second}$
&
$\SI{15.9}{\second}
$
&
$\pm 0.00$
&
$(1.00\text{--}1.00)$
& $\star$
&
$\SI{16.5}{\second}
$
&
$\pm 0.02$
&
$(0.97\text{--}1.04)$
&
\\
&
$\num[tight-spacing=true]{2.92e+08}$
&
$\num[tight-spacing=true]{2.54e+08}$
&
$\pm 0.00$
&
$(1.00\text{--}1.00)$
&
&
$\num[tight-spacing=true]{2.45e+08}$
&
$\pm 0.02$
&
$(0.96\text{--}1.05)$
& $\star$
\\[0.02cm]
frb30-15-2 & 
$\SI{1834.1}{\second}$
&
$\SI{32.3}{\second}
$
&
$\pm 0.00$
&
$(1.00\text{--}1.01)$
&
&
$\SI{31.6}{\second}
$
&
$\pm 0.05$
&
$(0.93\text{--}1.07)$
& $\star$
\\
&
$\num[tight-spacing=true]{5.57e+08}$
&
$\num[tight-spacing=true]{5.48e+08}$
&
$\pm 0.00$
&
$(1.00\text{--}1.00)$
&
&
$\num[tight-spacing=true]{5.08e+08}$
&
$\pm 0.06$
&
$(0.92\text{--}1.08)$
& $\star$
\\[0.02cm]
frb30-15-3 & 
$\SI{563.5}{\second}$
&
$\SI{7.9}{\second}
$
&
$\pm 0.01$
&
$(0.99\text{--}1.01)$
& $\star$
&
$\SI{8.5}{\second}
$
&
$\pm 0.03$
&
$(0.95\text{--}1.03)$
&
\\
&
$\num[tight-spacing=true]{1.67e+08}$
&
$\num[tight-spacing=true]{1.27e+08}$
&
$\pm 0.00$
&
$(1.00\text{--}1.01)$
&
&
$\num[tight-spacing=true]{1.27e+08}$
&
$\pm 0.03$
&
$(0.95\text{--}1.05)$
& $\star$
\\[0.02cm]
frb30-15-4 & 
$\SI{3136.8}{\second}$
&
$\SI{38.7}{\second}
$
&
$\pm 0.00$
&
$(1.00\text{--}1.00)$
& $\star$
&
$\SI{58.4}{\second}
$
&
$\pm 0.01$
&
$(0.98\text{--}1.02)$
&
\\
&
$\num[tight-spacing=true]{9.91e+08}$
&
$\num[tight-spacing=true]{6.79e+08}$
&
$\pm 0.00$
&
$(1.00\text{--}1.00)$
& $\star$
&
$\num[tight-spacing=true]{9.92e+08}$
&
$\pm 0.01$
&
$(0.98\text{--}1.02)$
&
\\[0.02cm]
frb30-15-5 & 
$\SI{913.3}{\second}$
&
$\SI{21.9}{\second}
$
&
$\pm 0.04$
&
$(0.92\text{--}1.06)$
&
&
$\SI{16.9}{\second}
$
&
$\pm 0.05$
&
$(0.93\text{--}1.09)$
& $\star$
\\
&
$\num[tight-spacing=true]{2.83e+08}$
&
$\num[tight-spacing=true]{3.80e+08}$
&
$\pm 0.05$
&
$(0.91\text{--}1.07)$
&
&
$\num[tight-spacing=true]{2.72e+08}$
&
$\pm 0.06$
&
$(0.92\text{--}1.10)$
& $\star$
\\ \bottomrule
        \end{tabular}
\end{table}

We now compare our new approach with an off-the-shelf randomised work stealing system, using the GCC
implementation of Intel's Cilk Plus extensions to C++. In \cref{table:parallel} we present
sequential and parallel runtimes for the larger solvable DIMACS instances, and the smallest group of
BHOSLIB instances (these are selected so that parallel runtimes are at least half a second---below
this point, measurements have considerable noise). Each parallel measurement is the average of ten
runs: we show both runtimes, and the number of recursive calls made. For parallel measurements, we
also give the standard deviation and the range of values observed, both expressed as a proportion.

For thirteen of the instances, resplitting both does less work and takes less time on average; in a
further six cases, resplitting takes less time, but does more work. Cilk Plus does both less work
and takes less time in six cases, and takes less time but does more work in one case. Additionally,
when Cilk Plus is faster, it is never by more than \SI{36}{\percent}; in one case, resplitting wins
by a factor of ten. (These results suggest that the assumption that each recursive call has the same
cost is not entirely safe---this is not surprising, since we should not expect all colourings to be
equally time-consuming. Additionally, our resplitting implementation counts requeued nodes multiple
times, whereas the Cilk Plus implementation does not, so resplitting will show a slightly higher
recursive call count when exactly the same amount of work is done.)

It may be possible to reduce the overheads incurred with Cilk Plus, through careful tuning.
However, the difference between the average runtimes is not the most striking part of the results.
When looking at the standard deviations and ranges, we see a more interesting picture: for many
instances, the runtimes for Cilk Plus vary substantially between repeat runs, whereas the
resplitting runtimes are much more consistent. (This is why we have not included Cilk Plus runtimes
on our speedup graphs: the speedups are too chaotic to give meaningful data.) This is particularly
visible with the ``brock'' and ``gen'' instances, which we have seen are very sensitive to the time
taken to find the solution: for ``gen400\_p0.9\_65'', our shortest observed runtime with Cilk Plus
was \SI{3477}{\second}, and our longest was \SI{12315}{\second}. By contrast, with resplitting, the
shortest runtime was \SI{4444}{\second} and the longest was \SI{4463}{\second}. We do not consider
the variability seen with Cilk Plus to be ideal behaviour.

Despite the lack of explicit diversity, we \emph{do} often see superlinear speedups with Cilk Plus.
We can explain this. The GCC Cilk Plus implementation tends to steal ``earliest created'' jobs
first: behind the scenes, each worker has a deque, and jobs are enqueued and processed by the worker
at one end (LIFO), but are stolen from the other (FIFO); victims for stealing are selected randomly.
The rationale is that this is less likely to introduce contention, and that earlier-created jobs are
likely to be larger \citep{McCool:2012}.  However, this has another benefit here: it introduces at
least some diversity early in the search. Assuming the $(1)$ subtree is non-trivial, the $(2)$
subtree will always be stolen immediately, and a third thread will then either steal $(1, 2)$ or
$(3)$, nondeterministically. This is an implementation detail, and not a specified behaviour, so it
could change with a new release of the compiler---this could have disastrous effects for parallel
branch and bound.

This accidental diversity also usually requires a higher number of threads before the incumbent can
be found quickly. With 16 threads as opposed to 64, Cilk Plus is slower for every instance except
``p\_hat1500-2'' (where it wins by less than \SI{1}{\percent}), due to it requiring more work to
find the solution.

\subsection{Other Approaches}

We have seen how this approach compares to that of \citet{Depolli:2013}: we improve the balance,
whilst retaining the diversity. Compared to an off-the-shelf work stealing implementation, we often
get better runtimes, and our performance is much more consistent between runs.

Compared to \citet{Xiang:2013}, we do not need to estimate up-front how large subtrees might be to
obtain balance, and we do not need to do any calibration.  Unfortunately Xiang et al.\ only
considered three of the DIMACS graphs, all from the same family.  None of these graphs have many
avoidable nodes, and so the effects we discuss here are not seen in their results---this explains
their consistently linear speedups (which are presented over a parallel algorithm, not a sequential
one). Their approach also does not preserve any sequential order, and only shares updated incumbents
when starting a new subproblem---when there are many avoidable nodes, this can cause a slowdown,
both in theory \citep{Trienekens:1990} and in practice \citep{McCreesh:2012}. It is thus unclear
what would happen with their approach on other DIMACS graph families, where scalability is not the
only concern.

Compared to Moisan et al.'s parallel discrepancy search \citep{Moisan:2013}, we are emphasising
early diversity, not total number of discrepancies---that is, we believe Harvey and Ginsberg's
second claim (that heuristics are weak early on) is important for this problem, not their first
(that the total number of wrong turns made is low). Another form of parallel limited discrepancy
search, in a constraint programming setting, is discussed by \citet{Michel:2009}. They report that
superlinear speedups are common for some problems when comparing parallel limited discrepancy search
to sequential depth first search. However, such an approach risks introducing a slowdown (which they
discuss), and again they consider the number of discrepancies rather than where those discrepancies
occur.

Compared to the work by Batsyn, Maslov et al., we are trying to obtain a strong incumbent earlier in
the search by using multiple paths through a single search tree, rather than a two stage process.
This means that all of our work done is potentially contributing to a proof of optimality. We also
do not need to select a pre-determined arbitrary time to run the first stage (Batsyn, Maslov et al.\
ran the heuristic for different amounts of time for different problems based upon pre-existing
knowledge of the probable size of a solution and the difficulty of the problem), so our approach
does not require special tuning before it can be used on ``unseen'' graphs. Note however that these
approaches are not mutually exclusive---for problems which are expected to be particularly
difficult, it would be possible to sacrifice one thread during early stages of our search process to
perform a search using a non-exact algorithm.

Another approach using multiple search trees is cooperating local search \citep{Clearwater:1991},
which has been demonstrated in a maximum clique context by \citet{Pullan:2011}. Here the aim is to
combine multiple non-exact algorithms in parallel in the hopes of gaining the strengths of each; no
attempt is made to prove optimality. This technique has also been used in graph colouring, where
\citet{Lewandowski:1996} observe that ``in general, having the processors work in parallel yields
better colorings faster than simply using multiple independent runs''.

This theme also shows up in the ManySAT parallel SAT solver \citep{Hamadi:2009}, where parallelism
is used to counter the sensitivity of SAT solvers to initial tuning parameters. Hamadi et al.\
remark that ``the performance of parallel solvers is usually better on SAT problems than on UNSAT
ones''. This is because multiple search trees may make it easier to find a solution, but they are
less helpful in proving that there is no solution (although things are not this simple for SAT,
where learned clauses are shared). In contrast, in cases where a good solution can be found quickly,
our approach allows all of the work done by every thread to contribute to a proof of
optimality---this is why we typically get at least near linear speedups.  Some SAT solvers also make
use of randomised restarts \citep{Gomes:1998}, again with the aim of avoiding heavy commitment to
any particular portion of the search space; the emphasis here is upon finding a solution to a
satisfiable instance more quickly.

Finally, one could use these experiments to support the view that \emph{sequential} maximum clique
algorithms should use some form of early discrepancy search. We would not dispute this, although we
note firstly that more complicated sequential search algorithms introduce overheads which cannot be
offset during the proof of optimality stage of search, and secondly that either way, every modern
processor has multiple cores and we should be making use of this \citep{Sutter:2005}. In any case,
doing this would not invalidate our approach---we would simply see some of our superlinear speedups
become linear speedups over a faster sequential baseline.

\section{Conclusion}

We have considered some of the design choices available when parallelising a state-of-the-art branch
and bound algorithm for the maximum clique problem. We have shown that the irregularity of
subproblem sizes can cause workload balance problems---this should not be surprising. But we also
saw that different parallel search orders will often produce substantially different speedups, to
the extent that balance is usually \emph{not} the deciding factor for performance.

This sheds new light on Lai and Sahni's claim that ``anomalous behaviour will be rarely witnessed in
practice'' \citep{Lai:1984}. Both the authors and Depolli et al.\ \emph{did} commonly encounter
superlinear speedups, but only because our parallel search orders encouraged this. We also saw
superlinear speedups when using Cilk Plus, but only because of an unintended effect of an
unspecified implementation detail. We used our understanding of the behaviour of heuristics to
provide an explanation: parallelism was introducing diversity into the search, avoiding a strong
commitment to weak early heuristic advice. We then showed how to preserve this diversity whilst
improving load balancing.

More generally, we have demonstrated that obtaining a parallel algorithm which seems to behave well
most of the time should only be the first step in the algorithm design process. For heuristics,
\citet{Hooker:1995} advocates a scientific approach to algorithm evaluation, rather than simple performance comparisons:

\begin{quote}``Based on one's insight into an algorithm, for instance,
one might expect good performance to depend on a certain characteristic. How to find out? Design a
controlled experiment that checks how the presence or absence of this characteristic affects
performance. Even better, build an explanatory mathematical model that captures the insight, as is
done routinely in other empirical sciences, and deduce from it precise consequences that can be put
to the test.''\end{quote}

Such an approach has been helpful here too.  By taking measurements inside search, and looking in
depth at individual results, we gained insight into how to improve a parallel algorithm. We
explained \emph{why} the authors and Depolli et al.\ sometimes saw superlinear speedups, and how to
preserve these speedups when modifying the parallel algorithm. We also showed why in certain cases
Depolli et al.\ had parallelism limits, and why Batsyn, Maslov et al.\ failed to obtain an
improvement by priming the search with a stronger incumbent for some problems.

We intend to apply this approach to other problems. In some ways, maximum clique is an ``ideal''
problem for this approach, and other problems may not have the same properties which could
complicate matters. We are aware of a potentially unfavourable interaction between this approach and
unbroken symmetries which must be addressed before some other hard problems may be considered. It is
also not obvious whether Harvey and Ginsberg's second claim (that heuristics are worst early in
search) will always hold---in cases where it does not, it may be necessary to use heuristic quality
information at runtime to direct parallel search order.

\section*{Acknowledgements}
The authors wish to thank Matja\v{z} Depolli for his comments and for making his source code
available, Jennifer Boyle and David Manlove for their comments, and Douglas MacFarlane and Joe
Sventek for their help with the hardware.

\bibliographystyle{ACM-Reference-Format-Journals}
\bibliography{early_splitting_in_max_clique}


\begin{thebibliography}{00}


\ifx \showCODEN    \undefined \def \showCODEN     #1{\unskip}     \fi
\ifx \showDOI      \undefined \def \showDOI       #1{{\tt DOI:}\penalty0{#1}\ }
  \fi
\ifx \showISBNx    \undefined \def \showISBNx     #1{\unskip}     \fi
\ifx \showISBNxiii \undefined \def \showISBNxiii  #1{\unskip}     \fi
\ifx \showISSN     \undefined \def \showISSN      #1{\unskip}     \fi
\ifx \showLCCN     \undefined \def \showLCCN      #1{\unskip}     \fi
\ifx \shownote     \undefined \def \shownote      #1{#1}          \fi
\ifx \showarticletitle \undefined \def \showarticletitle #1{#1}   \fi
\ifx \showURL      \undefined \def \showURL       #1{#1}          \fi

\bibitem[\protect\citeauthoryear{Bader, Hart, and Phillips}{Bader
  et~al\mbox{.}}{2005}]%
        {Bader:2005}
{David~A. Bader}, {William~E. Hart}, {and} {Cynthia~A. Phillips}. 2005.
\newblock \showarticletitle{Parallel Algorithm Design for Branch and Bound}.
\newblock In {\em Tutorials on Emerging Methodologies and Applications in
  Operations Research}, {HJ. G} (Ed.). International Series in Operations
  Research \& Management Science, Vol.~76. Springer New York, New York, NY,
  USA, 5--1--5--44.
\newblock
\showISBNx{978-0-387-22826-6}
\showDOI{%
\url{http://dx.doi.org/10.1007/0-387-22827-6_5}}


\bibitem[\protect\citeauthoryear{Batsyn, Goldengorin, Maslov, and
  Pardalos}{Batsyn et~al\mbox{.}}{2014}]%
        {Batsyn:2013}
{Mikhail Batsyn}, {Boris Goldengorin}, {Evgeny Maslov}, {and} {PanosM.
  Pardalos}. 2014.
\newblock \showarticletitle{Improvements to MCS algorithm for the maximum
  clique problem}.
\newblock {\em Journal of Combinatorial Optimization\/} {27}, 2 (2014),
  397--416.
\newblock
\showISSN{1382-6905}
\showDOI{%
\url{http://dx.doi.org/10.1007/s10878-012-9592-6}}


\bibitem[\protect\citeauthoryear{Bergman, Cire, Sabharwal, Samulowitz,
  Saraswat, and van Hoeve}{Bergman et~al\mbox{.}}{2014}]%
        {Bergman:2014}
{David Bergman}, {Andre~A. Cire}, {Ashish Sabharwal}, {Horst Samulowitz},
  {Vijay Saraswat}, {and} {Willem-Jan van Hoeve}. 2014.
\newblock \showarticletitle{Parallel Combinatorial Optimization with Decision
  Diagrams}.
\newblock In {\em Integration of AI and OR Techniques in Constraint
  Programming}, {Helmut Simonis} (Ed.). Lecture Notes in Computer Science, Vol.
  8451. Springer International Publishing, 351--367.
\newblock
\showISBNx{978-3-319-07045-2}
\showDOI{%
\url{http://dx.doi.org/10.1007/978-3-319-07046-9_25}}


\bibitem[\protect\citeauthoryear{Bomze, Budinich, Pardalos, and Pelillo}{Bomze
  et~al\mbox{.}}{1999}]%
        {Bomze:1999}
{Immanuel~M. Bomze}, {Marco Budinich}, {Panos~M. Pardalos}, {and} {Marcello
  Pelillo}. 1999.
\newblock \showarticletitle{The maximum clique problem}.
\newblock {\em Handbook of Combinatorial Optimization (Supplement Volume A)\/}
  {4} (1999), 1--74.
\newblock
\showURL{%
\url{http://citeseerx.ist.psu.edu/viewdoc/summary?doi=10.1.1.56.6221}}


\bibitem[\protect\citeauthoryear{Butenko and Wilhelm}{Butenko and
  Wilhelm}{2006}]%
        {Butenko:2006}
{Sergiy Butenko} {and} {Wilbert~E. Wilhelm}. 2006.
\newblock \showarticletitle{Clique-detection models in computational
  biochemistry and genomics}.
\newblock {\em European Journal of Operational Research\/} {173}, 1 (2006),
  1--17.
\newblock
\showDOI{%
\url{http://dx.doi.org/10.1016/j.ejor.2005.05.026}}


\bibitem[\protect\citeauthoryear{Caniou, Codognet, Diaz, and Abreu}{Caniou
  et~al\mbox{.}}{2011}]%
        {Caniou:2011}
{Yves Caniou}, {Philippe Codognet}, {Daniel Diaz}, {and} {Salvador Abreu}.
  2011.
\newblock \showarticletitle{Experiments in Parallel Constraint-Based Local
  Search}.
\newblock In {\em Evolutionary Computation in Combinatorial Optimization},
  {Peter Merz} {and} {Jin-Kao Hao} (Eds.). Lecture Notes in Computer Science,
  Vol. 6622. Springer Berlin Heidelberg, 96--107.
\newblock
\showISBNx{978-3-642-20363-3}
\showDOI{%
\url{http://dx.doi.org/10.1007/978-3-642-20364-0_9}}


\bibitem[\protect\citeauthoryear{Cheeseman, Kanefsky, and Taylor}{Cheeseman
  et~al\mbox{.}}{1991}]%
        {Cheeseman:1991}
{Peter Cheeseman}, {Bob Kanefsky}, {and} {William~M. Taylor}. 1991.
\newblock \showarticletitle{Where the really hard problems are}. In {\em
  Proceedings of the 12th international joint conference on Artificial
  intelligence - Volume 1} {\em (IJCAI'91)}. Morgan Kaufmann Publishers Inc.,
  San Francisco, CA, USA, 331--337.
\newblock
\showISBNx{1-55860-160-0}
\showURL{%
\url{http://dl.acm.org/citation.cfm?id=1631171.1631221}}


\bibitem[\protect\citeauthoryear{Clausen}{Clausen}{1997}]%
        {Clausen:1997}
{Jens Clausen}. 1997.
\newblock \showarticletitle{Parallel branch and bound---principles and personal
  experiences}.
\newblock In {\em Parallel Computing in Optimization}, {Athanasios Migdalas},
  {Panos~M. Pardalos}, {and} {Sverre Story} (Eds.). Kluwer Academic Publishers,
  Norwell, MA, USA, Chapter~7, 239--267.
\newblock
\showISBNx{0792345835}


\bibitem[\protect\citeauthoryear{Clearwater, Huberman, and Hogg}{Clearwater
  et~al\mbox{.}}{1991}]%
        {Clearwater:1991}
{Scott~H. Clearwater}, {Bernardo~A. Huberman}, {and} {Tad Hogg}. 1991.
\newblock \showarticletitle{Cooperative Solution of Constraint Satisfaction
  Problems}.
\newblock {\em Science\/} {254}, 5035 (1991), 1181--1183.
\newblock
\showDOI{%
\url{http://dx.doi.org/10.1126/science.254.5035.1181}}


\bibitem[\protect\citeauthoryear{Crainic, Le~Cun, and Roucairol}{Crainic
  et~al\mbox{.}}{2006}]%
        {Crainic:2006}
{Teodor~Gabriel Crainic}, {Bertrand Le~Cun}, {and} {Catherine Roucairol}. 2006.
\newblock {\em Parallel Branch-and-Bound Algorithms}.
\newblock John Wiley \& Sons, Inc., Hoboken, NJ, USA, 1--28.
\newblock
\showISBNx{9780470053928}
\showDOI{%
\url{http://dx.doi.org/10.1002/9780470053928.ch1}}


\bibitem[\protect\citeauthoryear{de~Bruin, Kindervater, and
  Trienekens}{de~Bruin et~al\mbox{.}}{1995}]%
        {deBruin:1995}
{A. de Bruin}, {G.A.P. Kindervater}, {and} {H.W.J.M. Trienekens}. 1995.
\newblock \showarticletitle{Asynchronous parallel branch and bound and
  anomalies}.
\newblock In {\em Parallel Algorithms for Irregularly Structured Problems},
  {Afonso Ferreira} {and} {José Rolim} (Eds.). Lecture Notes in Computer
  Science, Vol. 980. Springer Berlin Heidelberg, Berlin, Heidelberg, 363--377.
\newblock
\showISBNx{978-3-540-60321-4}
\showDOI{%
\url{http://dx.doi.org/10.1007/3-540-60321-2_29}}


\bibitem[\protect\citeauthoryear{Debroni, Eblen, Langston, Myrvold, Shor, and
  Weerapurage}{Debroni et~al\mbox{.}}{2011}]%
        {Debroni:2011}
{Jennifer Debroni}, {John~D. Eblen}, {Michael~A. Langston}, {Wendy Myrvold},
  {Peter Shor}, {and} {Dinesh Weerapurage}. 2011.
\newblock \showarticletitle{A Complete Resolution of the {Keller} Maximum
  Clique Problem}. In {\em Proceedings of the Twenty-second Annual ACM-SIAM
  Symposium on Discrete Algorithms} {\em (SODA '11)}. SIAM, 129--135.
\newblock
\showURL{%
\url{http://dl.acm.org/citation.cfm?id=2133036.2133047}}


\bibitem[\protect\citeauthoryear{Depolli, Konc, Rozman, Trobec, and
  Jane\v{z}i\v{c}}{Depolli et~al\mbox{.}}{2013}]%
        {Depolli:2013}
{Matja\v{z} Depolli}, {Janez Konc}, {Kati Rozman}, {Roman Trobec}, {and}
  {Du\v{s}anka Jane\v{z}i\v{c}}. 2013.
\newblock \showarticletitle{Exact Parallel Maximum Clique Algorithm for General
  and Protein Graphs}.
\newblock {\em Journal of Chemical Information and Modeling\/} {53}, 9 (2013),
  2217--2228.
\newblock
\showDOI{%
\url{http://dx.doi.org/10.1021/ci4002525}}


\bibitem[\protect\citeauthoryear{Eblen, Phillips, Rogers, and Langston}{Eblen
  et~al\mbox{.}}{2011}]%
        {Eblen:2011}
{John~D. Eblen}, {Charles~A. Phillips}, {Gary~L. Rogers}, {and} {Michael~A.
  Langston}. 2011.
\newblock \showarticletitle{The Maximum Clique Enumeration Problem: Algorithms,
  Applications and Implementations}.
\newblock In {\em Bioinformatics Research and Applications}, {Jianer Chen},
  {Jianxin Wang}, {and} {Alexander Zelikovsky} (Eds.). Lecture Notes in
  Computer Science, Vol. 6674. Springer Berlin Heidelberg, Berlin, Heidelberg,
  306--319.
\newblock
\showISBNx{978-3-642-21259-8}
\showDOI{%
\url{http://dx.doi.org/10.1007/978-3-642-21260-4_30}}


\bibitem[\protect\citeauthoryear{Fischetti, Monaci, and Salvagnin}{Fischetti
  et~al\mbox{.}}{2014}]%
        {Fischetti:2014}
{Matteo Fischetti}, {Michele Monaci}, {and} {Domenico Salvagnin}. 2014.
\newblock \showarticletitle{Self-splitting of Workload in Parallel
  Computation}.
\newblock In {\em Integration of AI and OR Techniques in Constraint
  Programming}, {Helmut Simonis} (Ed.). Lecture Notes in Computer Science, Vol.
  8451. Springer International Publishing, 394--404.
\newblock
\showISBNx{978-3-319-07045-2}
\showDOI{%
\url{http://dx.doi.org/10.1007/978-3-319-07046-9_28}}


\bibitem[\protect\citeauthoryear{Garey and Johnson}{Garey and Johnson}{1990}]%
        {Garey:1990}
{Michael~R. Garey} {and} {David~S. Johnson}. 1990.
\newblock {\em Computers and Intractability; A Guide to the Theory of
  {NP}-Completeness}.
\newblock W. H. Freeman \& Co., New York, NY, USA.
\newblock
\showISBNx{0716710455}


\bibitem[\protect\citeauthoryear{Gomes, Selman, and Kautz}{Gomes
  et~al\mbox{.}}{1998}]%
        {Gomes:1998}
{Carla~P. Gomes}, {Bart Selman}, {and} {Henry Kautz}. 1998.
\newblock \showarticletitle{Boosting Combinatorial Search Through
  Randomization}. In {\em Proceedings of the Fifteenth National/Tenth
  Conference on Artificial Intelligence/Innovative Applications of Artificial
  Intelligence} {\em (AAAI '98/IAAI '98)}. American Association for Artificial
  Intelligence, Palo Alto, CA, USA, 431--437.
\newblock
\showISBNx{0-262-51098-7}
\showURL{%
\url{http://dl.acm.org/citation.cfm?id=295240.295710}}


\bibitem[\protect\citeauthoryear{Hamadi, Jabbour, and Sais}{Hamadi
  et~al\mbox{.}}{2009}]%
        {Hamadi:2009}
{Youssef Hamadi}, {Said Jabbour}, {and} {Lakhdar Sais}. 2009.
\newblock \showarticletitle{{ManySAT}: a parallel {SAT} solver}.
\newblock {\em Journal on Satisfiability, Boolean Modeling and Computation
  (JSAT)\/}  {6} (2009), 245--262.
\newblock


\bibitem[\protect\citeauthoryear{Harvey and Ginsberg}{Harvey and
  Ginsberg}{1995}]%
        {Harvey:1995}
{William~D. Harvey} {and} {Matthew~L. Ginsberg}. 1995.
\newblock \showarticletitle{Limited Discrepancy Search}. In {\em IJCAI (1)}.
  Morgan Kaufmann, San Francisco, CA, USA, 607--615.
\newblock


\bibitem[\protect\citeauthoryear{Hooker}{Hooker}{1995}]%
        {Hooker:1995}
{J.N. Hooker}. 1995.
\newblock \showarticletitle{Testing heuristics: We have it all wrong}.
\newblock {\em Journal of Heuristics\/} {1}, 1 (1995), 33--42.
\newblock
\showISSN{1381-1231}
\showDOI{%
\url{http://dx.doi.org/10.1007/BF02430364}}


\bibitem[\protect\citeauthoryear{Korf}{Korf}{1996}]%
        {Korf:1996}
{Richard~E. Korf}. 1996.
\newblock \showarticletitle{Improved Limited Discrepancy Search.}. In {\em
  AAAI/IAAI, Vol. 1}. AAAI Press, Palo Alto, CA, USA, 286--291.
\newblock


\bibitem[\protect\citeauthoryear{Lai and Sahni}{Lai and Sahni}{1984}]%
        {Lai:1984}
{Ten-Hwang Lai} {and} {Sartaj Sahni}. 1984.
\newblock \showarticletitle{Anomalies in parallel branch-and-bound algorithms}.
\newblock {\it Commun. ACM} {27}, 6 (1984), 594--602.
\newblock


\bibitem[\protect\citeauthoryear{Langer, Venkataraman, Palekar, and
  Kale}{Langer et~al\mbox{.}}{2013}]%
        {Langer:2013}
{A Langer}, {R. Venkataraman}, {U. Palekar}, {and} {L.V. Kale}. 2013.
\newblock \showarticletitle{Parallel branch-and-bound for two-stage stochastic
  integer optimization}. In {\em High Performance Computing (HiPC), 2013 20th
  International Conference on}. 266--275.
\newblock
\showDOI{%
\url{http://dx.doi.org/10.1109/HiPC.2013.6799130}}


\bibitem[\protect\citeauthoryear{Leroy, Mezmaz, Melab, and Tuyttens}{Leroy
  et~al\mbox{.}}{2014}]%
        {Leroy:2014}
{Rudi Leroy}, {Mohand Mezmaz}, {Nouredine Melab}, {and} {Daniel Tuyttens}.
  2014.
\newblock \showarticletitle{Work Stealing Strategies For Multi-Core Parallel
  Branch-and-Bound Algorithm Using Factorial Number System}. In {\em
  Proceedings of Programming Models and Applications on Multicores and
  Manycores} {\em (PMAM'14)}. ACM, New York, NY, USA, Article 111, 9 pages.
\newblock
\showISBNx{978-1-4503-2657-5}
\showDOI{%
\url{http://dx.doi.org/10.1145/2560683.2560694}}


\bibitem[\protect\citeauthoryear{Lewandowski and Condon}{Lewandowski and
  Condon}{1996}]%
        {Lewandowski:1996}
{Gary Lewandowski} {and} {Anne Condon}. 1996.
\newblock \showarticletitle{Experiments with parallel graph coloring heuristics
  and applications of graph coloring}.
\newblock {\em {DIMACS} series in discrete mathematics and theoretical computer
  science\/}  {26} (1996), 309--334.
\newblock


\bibitem[\protect\citeauthoryear{Li, Zhu, Many\`{a}, and Simon}{Li
  et~al\mbox{.}}{2011}]%
        {Li:2011}
{Chu-Min Li}, {Zhu Zhu}, {Felip Many\`{a}}, {and} {Laurent Simon}. 2011.
\newblock \showarticletitle{Minimum Satisfiability and Its Applications}. In
  {\em Proceedings of the Twenty-Second International Joint Conference on
  Artificial Intelligence - Volume Volume One} {\em (IJCAI'11)}. AAAI Press,
  Palo Alto, CA, USA, 605--610.
\newblock
\showISBNx{978-1-57735-513-7}
\showDOI{%
\url{http://dx.doi.org/10.5591/978-1-57735-516-8/IJCAI11-108}}


\bibitem[\protect\citeauthoryear{Li and Wah}{Li and Wah}{1986}]%
        {Li:1985}
{Guo-Jie Li} {and} {B. Wah}. 1986.
\newblock \showarticletitle{Coping with Anomalies in Parallel Branch-and-Bound
  Algorithms}.
\newblock {\em Computers, IEEE Transactions on\/} {C-35}, 6 (1986), 568--573.
\newblock
\showISSN{0018-9340}
\showDOI{%
\url{http://dx.doi.org/10.1109/TC.1986.5009434}}


\bibitem[\protect\citeauthoryear{Maslov, Batsyn, and Pardalos}{Maslov
  et~al\mbox{.}}{2013}]%
        {Maslov:2013}
{Evgeny Maslov}, {Mikhail Batsyn}, {and} {Panos~M. Pardalos}. 2013.
\newblock \showarticletitle{Speeding up {MCS} Algorithm for the Maximum Clique
  Problem with {ILS} Heuristic and Other Enhancements}.
\newblock In {\em Models, Algorithms, and Technologies for Network Analysis},
  {Boris~I. Goldengorin}, {Valery~A. Kalyagin}, {and} {Panos~M. Pardalos}
  (Eds.). Springer Proceedings in Mathematics \& Statistics, Vol.~59. Springer
  New York, New York, NY, USA, 93--99.
\newblock
\showISBNx{978-1-4614-8587-2}
\showDOI{%
\url{http://dx.doi.org/10.1007/978-1-4614-8588-9_7}}


\bibitem[\protect\citeauthoryear{McCool, Reinders, and Robison}{McCool
  et~al\mbox{.}}{2012}]%
        {McCool:2012}
{Michael McCool}, {James Reinders}, {and} {Arch Robison}. 2012.
\newblock {\em Structured parallel programming: patterns for efficient
  computation}.
\newblock Elsevier.
\newblock


\bibitem[\protect\citeauthoryear{McCreesh and Prosser}{McCreesh and
  Prosser}{2012}]%
        {McCreesh:2012}
{Ciaran McCreesh} {and} {Patrick Prosser}. 2012.
\newblock \showarticletitle{Distributing an Exact Algorithm for Maximum Clique:
  maximising the costup}.
\newblock {\em CoRR\/}  {abs/1209.4560} (2012).
\newblock


\bibitem[\protect\citeauthoryear{McCreesh and Prosser}{McCreesh and
  Prosser}{2013a}]%
        {McCreesh:2013.misleading}
{Ciaran McCreesh} {and} {Patrick Prosser}. 2013a.
\newblock \showarticletitle{Greedy Graph Colouring is a Misleading Heuristic}.
\newblock {\em CoRR\/}  {abs/1310.7741} (2013).
\newblock


\bibitem[\protect\citeauthoryear{McCreesh and Prosser}{McCreesh and
  Prosser}{2013b}]%
        {McCreesh:2013}
{Ciaran McCreesh} {and} {Patrick Prosser}. 2013b.
\newblock \showarticletitle{Multi-Threading a State-of-the-Art Maximum Clique
  Algorithm}.
\newblock {\em Algorithms\/} {6}, 4 (2013), 618--635.
\newblock
\showISSN{1999-4893}
\showDOI{%
\url{http://dx.doi.org/10.3390/a6040618}}


\bibitem[\protect\citeauthoryear{Michel, See, and Van~Hentenryck}{Michel
  et~al\mbox{.}}{2009}]%
        {Michel:2009}
{Laurent Michel}, {Andrew See}, {and} {Pascal Van~Hentenryck}. 2009.
\newblock \showarticletitle{Transparent Parallelization of Constraint
  Programming}.
\newblock {\em INFORMS Journal on Computing\/} {21}, 3 (2009), 363--382.
\newblock
\showDOI{%
\url{http://dx.doi.org/10.1287/ijoc.1080.0313}}


\bibitem[\protect\citeauthoryear{Moisan, Gaudreault, and Quimper}{Moisan
  et~al\mbox{.}}{2013}]%
        {Moisan:2013}
{Thierry Moisan}, {Jonathan Gaudreault}, {and} {Claude-Guy Quimper}. 2013.
\newblock \showarticletitle{Parallel Discrepancy-Based Search}.
\newblock In {\em Principles and Practice of Constraint Programming},
  {Christian Schulte} (Ed.). Lecture Notes in Computer Science, Vol. 8124.
  Springer Berlin Heidelberg, Berlin, Heidelberg, 30--46.
\newblock
\showISBNx{978-3-642-40626-3}
\showDOI{%
\url{http://dx.doi.org/10.1007/978-3-642-40627-0_6}}


\bibitem[\protect\citeauthoryear{Pardalos, Rappe, Mauricio, and
  Resende}{Pardalos et~al\mbox{.}}{1998}]%
        {Pardalos:1998}
{Panos~M. Pardalos}, {Jonas Rappe}, {Mauricio}, {and} {Mauricio~G.C. Resende}.
  1998.
\newblock \showarticletitle{An Exact Parallel Algorithm For The Maximum Clique
  Problem}. In {\em In High Performance and Software in Nonlinear
  Optimization}. Kluwer Academic Publishers, Dordrecht, the Netherlands,
  279--300.
\newblock


\bibitem[\protect\citeauthoryear{Poldner and Kuchen}{Poldner and
  Kuchen}{2008}]%
        {Poldner:2008}
{Michael Poldner} {and} {Herbert Kuchen}. 2008.
\newblock \showarticletitle{Algorithmic Skeletons for Branch and Bound}.
\newblock In {\em Software and Data Technologies}, {Joaquim Filipe}, {Boris
  Shishkov}, {and} {Markus Helfert} (Eds.). Communications in Computer and
  Information Science, Vol.~10. Springer Berlin Heidelberg, 204--219.
\newblock
\showISBNx{978-3-540-70619-9}
\showDOI{%
\url{http://dx.doi.org/10.1007/978-3-540-70621-2_17}}


\bibitem[\protect\citeauthoryear{Prosser}{Prosser}{2012}]%
        {Prosser:2012}
{Patrick Prosser}. 2012.
\newblock \showarticletitle{Exact algorithms for maximum clique: a
  computational study}.
\newblock {\em Algorithms\/} {5}, 4 (2012), 545--587.
\newblock
\showISSN{1999-4893}
\showDOI{%
\url{http://dx.doi.org/10.3390/a5040545}}


\bibitem[\protect\citeauthoryear{Prosser and Unsworth}{Prosser and
  Unsworth}{2011}]%
        {Prosser:2011}
{Patrick Prosser} {and} {Chris Unsworth}. 2011.
\newblock \showarticletitle{Limited discrepancy search revisited}.
\newblock {\em Journal of Experimental Algorithmics (JEA)\/}  {16} (2011),
  1--6.
\newblock


\bibitem[\protect\citeauthoryear{Pullan, Mascia, and Brunato}{Pullan
  et~al\mbox{.}}{2011}]%
        {Pullan:2011}
{Wayne Pullan}, {Franco Mascia}, {and} {Mauro Brunato}. 2011.
\newblock \showarticletitle{Cooperating local search for the maximum clique
  problem}.
\newblock {\em Journal of heuristics\/} {17}, 2 (2011), 181--199.
\newblock


\bibitem[\protect\citeauthoryear{Rayward-Smith, Rush, and
  McKeown}{Rayward-Smith et~al\mbox{.}}{1993}]%
        {RaywardSmith:1993}
{V.J. Rayward-Smith}, {S.A. Rush}, {and} {G.P. McKeown}. 1993.
\newblock \showarticletitle{Efficiency considerations in the implementation of
  parallel branch-and-bound}.
\newblock {\em Annals of Operations Research\/} {43}, 2 (1993), 123--145.
\newblock
\showISSN{0254-5330}
\showDOI{%
\url{http://dx.doi.org/10.1007/BF02024489}}


\bibitem[\protect\citeauthoryear{R{\'e}gin}{R{\'e}gin}{2003}]%
        {Regin:2003}
{Jean-Charles R{\'e}gin}. 2003.
\newblock \showarticletitle{Using Constraint Programming to Solve the Maximum
  Clique Problem}.
\newblock In {\em Principles and Practice of Constraint Programming – {CP}
  2003}, {Francesca Rossi} (Ed.). Lecture Notes in Computer Science, Vol. 2833.
  Springer Berlin Heidelberg, Berlin, Heidelberg, 634--648.
\newblock
\showISBNx{978-3-540-20202-8}
\showDOI{%
\url{http://dx.doi.org/10.1007/978-3-540-45193-8_43}}


\bibitem[\protect\citeauthoryear{R\'{e}gin, Rezgui, and Malapert}{R\'{e}gin
  et~al\mbox{.}}{2013}]%
        {Regin:2013}
{Jean-Charles R\'{e}gin}, {Mohamed Rezgui}, {and} {Arnaud Malapert}. 2013.
\newblock \showarticletitle{Embarrassingly Parallel Search}.
\newblock In {\em Principles and Practice of Constraint Programming},
  {Christian Schulte} (Ed.). Lecture Notes in Computer Science, Vol. 8124.
  Springer Berlin Heidelberg, Berlin, Heidelberg, 596--610.
\newblock
\showISBNx{978-3-642-40626-3}
\showDOI{%
\url{http://dx.doi.org/10.1007/978-3-642-40627-0_45}}


\bibitem[\protect\citeauthoryear{Regula and Lantos}{Regula and Lantos}{2013}]%
        {Regula:2013}
{Gergely Regula} {and} {B{\'e}la Lantos}. 2013.
\newblock \showarticletitle{Formation Control of Quadrotor Helicopters with
  Guaranteed Collision Avoidance via Safe Path}.
\newblock {\em Electrical Engineering and Computer Science\/} {56}, 4 (2013),
  113--124.
\newblock


\bibitem[\protect\citeauthoryear{Rossi, Gleich, Gebremedhin, and Patwary}{Rossi
  et~al\mbox{.}}{2013}]%
        {Rossi:2013}
{Ryan~A. Rossi}, {David~F. Gleich}, {Assefaw~Hadish Gebremedhin}, {and} {Md.
  Mostofa~Ali Patwary}. 2013.
\newblock \showarticletitle{A Fast Parallel Maximum Clique Algorithm for Large
  Sparse Graphs and Temporal Strong Components}.
\newblock {\em CoRR\/}  {abs/1302.6256} (2013).
\newblock


\bibitem[\protect\citeauthoryear{San~Segundo, Lopez, and Batsyn}{San~Segundo
  et~al\mbox{.}}{2014}]%
        {SanSegundo:2014}
{Pablo San~Segundo}, {Alvaro Lopez}, {and} {Mikhail Batsyn}. 2014.
\newblock \showarticletitle{Initial Sorting of Vertices in the Maximum Clique
  Problem Reviewed}.
\newblock In {\em Learning and Intelligent Optimization}, {Panos~M. Pardalos},
  {Mauricio~G.C. Resende}, {Chrysafis Vogiatzis}, {and} {Jose~L. Walteros}
  (Eds.). Springer International Publishing, 111--120.
\newblock
\showISBNx{978-3-319-09583-7}
\showDOI{%
\url{http://dx.doi.org/10.1007/978-3-319-09584-4_12}}


\bibitem[\protect\citeauthoryear{San~Segundo, Matia, Rodriguez-Losada, and
  Hernando}{San~Segundo et~al\mbox{.}}{2013}]%
        {SanSegundo:2011b}
{Pablo San~Segundo}, {Fernando Matia}, {Diego Rodriguez-Losada}, {and} {Miguel
  Hernando}. 2013.
\newblock \showarticletitle{An improved bit parallel exact maximum clique
  algorithm}.
\newblock {\em Optimization Letters\/} {7}, 3 (2013), 467--479.
\newblock
\showISSN{1862-4472}
\showDOI{%
\url{http://dx.doi.org/10.1007/s11590-011-0431-y}}


\bibitem[\protect\citeauthoryear{San~Segundo, Rodr\'{\i}guez-Losada, and
  Jim{\'e}nez}{San~Segundo et~al\mbox{.}}{2011}]%
        {SanSegundo:2011}
{Pablo San~Segundo}, {Diego Rodr\'{\i}guez-Losada}, {and} {Agust\'{\i}n
  Jim{\'e}nez}. 2011.
\newblock \showarticletitle{An exact bit-parallel algorithm for the maximum
  clique problem}.
\newblock {\em Comput. Oper. Res.\/} {38}, 2 (Feb. 2011), 571--581.
\newblock
\showISSN{0305-0548}
\showDOI{%
\url{http://dx.doi.org/10.1016/j.cor.2010.07.019}}


\bibitem[\protect\citeauthoryear{Sanders}{Sanders}{1995}]%
        {Sanders:1995}
{Peter Sanders}. 1995.
\newblock \showarticletitle{Better algorithms for parallel backtracking}.
\newblock In {\em Parallel Algorithms for Irregularly Structured Problems},
  {Afonso Ferreira} {and} {José Rolim} (Eds.). Lecture Notes in Computer
  Science, Vol. 980. Springer Berlin Heidelberg, Berlin, Heidelberg, 333--347.
\newblock
\showISBNx{978-3-540-60321-4}
\showDOI{%
\url{http://dx.doi.org/10.1007/3-540-60321-2_27}}


\bibitem[\protect\citeauthoryear{Sutter}{Sutter}{2005}]%
        {Sutter:2005}
{Herb Sutter}. 2005.
\newblock \showarticletitle{The Free Lunch Is Over: A Fundamental Turn Toward
  Concurrency in Software}.
\newblock {\em Dr. Dobb's Journal\/} {30}, 3 (2005).
\newblock


\bibitem[\protect\citeauthoryear{Tomita and Kameda}{Tomita and Kameda}{2007}]%
        {Tomita:2007}
{E. Tomita} {and} {T. Kameda}. 2007.
\newblock \showarticletitle{An efficient branch-and-bound algorithm for finding
  a maximum clique with computational experiments}.
\newblock {\em Journal of Global Optimization\/} {37}, 1 (2007), 95--111.
\newblock


\bibitem[\protect\citeauthoryear{Tomita and Seki}{Tomita and Seki}{2003}]%
        {Tomita:2003}
{Etsuji Tomita} {and} {Tomokazu Seki}. 2003.
\newblock \showarticletitle{An efficient branch-and-bound algorithm for finding
  a maximum clique}. In {\em Proceedings of the 4th international conference on
  Discrete mathematics and theoretical computer science} {\em (DMTCS'03)}.
  Springer-Verlag, Berlin, Heidelberg, 278--289.
\newblock
\showISBNx{3-540-40505-4}
\showURL{%
\url{http://dl.acm.org/citation.cfm?id=1783712.1783736}}


\bibitem[\protect\citeauthoryear{Tomita, Sutani, Higashi, Takahashi, and
  Wakatsuki}{Tomita et~al\mbox{.}}{2010}]%
        {Tomita:2010}
{Etsuji Tomita}, {Yoichi Sutani}, {Takanori Higashi}, {Shinya Takahashi}, {and}
  {Mitsuo Wakatsuki}. 2010.
\newblock \showarticletitle{A Simple and Faster Branch-and-Bound Algorithm for
  Finding a Maximum Clique}.
\newblock In {\em WALCOM: Algorithms and Computation}, {Md.Saidur Rahman} {and}
  {Satoshi Fujita} (Eds.). Lecture Notes in Computer Science, Vol. 5942.
  Springer Berlin Heidelberg, Berlin, Heidelberg, 191--203.
\newblock
\showISBNx{978-3-642-11439-7}
\showDOI{%
\url{http://dx.doi.org/10.1007/978-3-642-11440-3_18}}


\bibitem[\protect\citeauthoryear{Trienekens}{Trienekens}{1990}]%
        {Trienekens:1990}
{Harry~W.J.M. Trienekens}. 1990.
\newblock {\em Parallel Branch and Bound Algorithms}.
\newblock Ph.D. Dissertation. Erasmus University Rotterdam.
\newblock


\bibitem[\protect\citeauthoryear{Walsh}{Walsh}{1997}]%
        {Walsh:1997}
{Toby Walsh}. 1997.
\newblock \showarticletitle{Depth-bounded Discrepancy Search}. In {\em IJCAI}.
  Morgan Kaufmann, San Francisco, CA, USA, 1388--1395.
\newblock


\bibitem[\protect\citeauthoryear{Xiang, Guo, and Aboulnaga}{Xiang
  et~al\mbox{.}}{2013}]%
        {Xiang:2013}
{Jingen Xiang}, {Cong Guo}, {and} {A. Aboulnaga}. 2013.
\newblock \showarticletitle{Scalable maximum clique computation using
  MapReduce}. In {\em Data Engineering (ICDE), 2013 IEEE 29th International
  Conference on}. 74--85.
\newblock
\showISSN{1063-6382}
\showDOI{%
\url{http://dx.doi.org/10.1109/ICDE.2013.6544815}}


\end{thebibliography}

\end{document}